\documentclass[aps,prd,amssymb,eqsecnum,showpacs]{revtex4-1}
\usepackage{graphicx}
\usepackage{psfrag}
\usepackage[usenames]{color}
\usepackage{amsmath}
\newcommand{\Lie}[1]{\mathcal{L}_\mathbf{#1}}

\begin{document}

\title{The Merger of Small and Large Black Holes}

\author{P. M{\" o}sta${}^{1,2}$, L. Andersson${}^{1}$, J. Metzger${}^{1,3}$,
B. Szil\'{a}gyi${}^{1,2}$,
            and J. Winicour${}^{1,4}$
       }
\affiliation{
${}^{1}$ Max-Planck-Institut f\" ur
         Gravitationsphysik, Albert-Einstein-Institut, 
	  14476 Golm, Germany\\
${}^{2}$ TAPIR, California Institute of Technology \\
               Pasadena, CA 91125, USA \\
${}^{3}$ Institit f{\" u}r Mathematik, Universit{\" a}t Potsdam \\
              14469 Potsdam, Germany \\               
${}^{4}$ Department of Physics and Astronomy \\
         University of Pittsburgh, Pittsburgh, PA 15260, USA 
	 }

\begin{abstract}

We present simulations of binary black holes mergers in which, after the common outer horizon has
formed, the marginally outer trapped surfaces (MOTSs) corresponding to the individual black holes
continue to approach and eventually penetrate each other. This has very interesting consequences
according to recent results in the theory of MOTSs.
Uniqueness and stability theorems imply that two
MOTSs which touch with a common outer normal must be identical. This suggests
a possible dramatic consequence of the collision between a small and large black hole.
If the penetration were to continue to completion then the two MOTSs would have
to coalesce, by some combination of the small one growing and the big one
shrinking.  Here we explore the relationship
between theory and numerical simulations, in which
a small black hole has halfway penetrated a large one.

\end{abstract}


\maketitle

\section{Introduction}

An unexpected feature has been observed in the simulation of equal mass binary black holes
following their inspiral and merger~\cite{pen}. After formation of a common
apparent horizon, as the two marginally outer trapped surfaces (MOTSs)
corresponding to the individual black holes continued to approach, they
eventually touched and then penetrated each other. This penetration was
surprising because it had not been considered in theoretical discussions and had
not been observed in prior simulations of binary black hole mergers. In
retrospect, it is tempting to speculate in some heuristic sense that a small
black hole should enter a very large black hole with hardly any notice of its
presence. In fact, it has been been conjectured on the basis of the equivalence
principle that a very small black hole should, in some appropriate sense, fall
into the large one along a geodesic. However, such a perturbative picture is
unreliable in the interior of the event horizon surrounding the two black holes,
where the MOTSs exist. Here we use an independent evolution code to first confirm
that the individual MOTSs do penetrate following a binary black hole merger and
we analyze some of the highly interesting features revealed by the simulations.

The simulations which first demonstrated the penetration of the individual MOTSs 
were carried out with a code based upon Fock's treatment~\cite{fock}
of the harmonic formulation in terms of the densitized metric $\sqrt{-g}g^{\mu\nu}$,
i.e., the PITT Abigel code which had been developed for the purpose
of studying outer boundary
conditions~\cite{bsw,bkw}. By incorporating adaptive mesh refinement and a
horizon tracker available in the CACTUS toolkit~\cite{cactus_web}, the code was
further developed to simulate a binary black hole inspiral using excision to
deal with the internal singularities. The simulations presented in the present
paper are based upon an independent harmonic code using standard $3+1$
variables and treating the singularities via punctures~\cite{punct} rather than
excision~\cite{excis}. We describe the computational methods in
Sec.~\ref{sec:compmeth}.

The seminal work of Hayward~\cite{hayw} and Ashtekar and Krishnan~\cite{ashkr}
has led to a rich mathematical theory of the dynamical horizons traced out
by the evolution of MOTSs, as reviewed in~\cite{ashkrlr}. This theory has
provided insight into both the classical and quantum properties of black holes.
In particular, the subject has strong bearing on numerical relativity because a
MOTS can be identified quasilocally by the  vanishing expansion of its outward
directed null rays, whereas the identification of an event horizon requires
global information which is not available at the early stages of a simulation.
Thus MOTSs play a central role in the preparation of initial data sets for black
holes and in tracking their subsequent evolution. A numerical study of MOTSs has
been carried out~\cite{schkr,schkrb}, which confirmed the main features expected
from the theory regarding the early stages of a binary inspiral. Recently, there
has been substantial new theoretical development centered around the uniqueness
and stability of
MOTSs~\cite{Andersson-Mars-Simon:2005,Andersson-Mars-Simon:2008,AMMS:2009,Ashthekar-Galloway:2005}.
This recent theory applies to the general case of the marginally outer trapped tube (MOTT)
traced out by a MOTS, which need not be spacelike as in the special case
of a dynamical horizon. It has important bearing on how
MOTSs approach each other and penetrate. We review the main mathematical results
and their relevance to the binary black hole problem in Sec.~\ref{sec:theory}.

The simulations presented in Sec.~\ref{sec:simulations} confirm the results
observed in~\cite{pen} and significantly extend the degree of penetration of the
MOTSs. There are five distinct stages as the MOTSs approach and penetrate.

\begin{itemize}

\item The large separation of the individual MOTSs. The properties of this
stage are mainly determined by the choice of binary black hole initial data.

\item Formation of a common apparent horizon as the MOTSs approach.

\item The moment of external osculation of the two MOTSs as they touch.

\item The penetration of the two MOTSs.

\item The ultimate fate of the individual MOTSs.
 
\end{itemize}

The original simulation showing the penetration of two inspiraling MOTSs~\cite{pen} and
the numerical study of MOTSs in the early stages of a binary
black hole~\cite{schkrb} were confined to the case of equal mass black holes. Here we
consider the unequal mass case, with the goal of shedding light on what happens
in the extreme mass ratio regime. Because of the computational cost in resolving
different length scales, we limit ourselves to  the case of a mass ratio
$m_{small}/m_{large}=1/4$. In order to better understand the geometry of the
merger, we also restrict ourselves to the simplest case of a head-on collision.
Already in this case, there are quite complicated geometrical effects.

Our simulations are based upon time symmetric initial data for the two black
holes. The time symmetry introduces some non-intuitive features of the initial
MOTSs, particularly when they are close together, which emphasize the importance
of looking at their invariant geometrical properties rather than their
coordinate description. We repeat, for the unequal mass case, the simulation of
the head-on collision carried out by Krishnan and Schnetter~\cite{schkrb} for
equal mass black holes. They simulated this equal mass case with 
a code based upon a $3+1$ formulation of Einstein's equations (the AEI
BSSN formulation), which is quite different from the harmonic formulation used
here. As a check on our code, we find qualitative agreement with the results
of~\cite{schkrb} for the early stages before the individual MOTSs touch. We
then continue the simulation of the head-on collision to the later stage where the individual
MOTSs penetrate, which could not be treated in~\cite{schkrb}. 

The original simulations of the penetration in~\cite{pen} were limited by the
excision region inside the individual MOTSs. The present code is able
to track the penetration to  a much further stage, essentially until
the small MOTS has penetrated halfway into the large MOTS. At that stage its
interior puncture is close to the surface of the large MOTS and the horizon finder
is not able to track it. The simulation of the penetration is validated by
the convergence measurements
given in Sec.~\ref{sec:convergence}.

We demonstrate by numerical simulation that before the individual MOTSs make
contact a common outer apparent horizon forms, in accord with the theory
described in Sec.~\ref{sec:theory}. When the two MOTSs first touch, the theory
also predicts that their mean surface curvatures must agree at the point of
osculation. Consistent with this theory, our simulations show that the small MOTS
creates a strong distortion of the large MOTS
at the contact point which causes the mean curvature of the large MOTS
to grow rapidly at the point of contact,
i.e. its curvature radius rapidly shrinks. On the other hand, at this stage
the large MOTS has only a small effect on the small one. We
track this growth of mean curvature of the large MOTS as the penetration
proceeds. In a scenario where the black holes are produced by collapsing
matter, without any puncture, one might expect the penetration to continue to
the point where the small MOTS completely penetrates the large one. If this
were to occur, then a dramatic consequence of the underlying uniqueness theorems
for MOTSs would ensue. At the time where the back of the small MOTS enters the
front of the large one, the theorems imply that the two MOTSs must identically
coalesce, either through shrinkage of the large MOTS or growth of the small
MOTS. Although our simulations cannot proceed to this stage, they provide some
glimpse of how it might proceed.

\section{Computational method}
\label{sec:compmeth}

\subsection{The evolution system}

We employ an evolution system based upon the 3+1 formulation of
Einstein equations in generalized harmonic coordinates $x^\mu=(t,x^i)$
as described by Friedrich and Rendall~\cite{friedren}.
(We make minor alterations
in the notation of~\cite{friedren} to conform to standard usage in numerical relativity.)
This 3+1 foliation gives rise to the metric decomposition
\begin{equation} 
       g_{\mu\nu}= -n_\mu n_\nu +h_{\mu\nu} 
\end{equation} 
where 
\begin{equation}
      n_\mu=-\alpha \partial_\mu t , \quad \alpha =\frac{1}{ \sqrt {-g^{tt}}}
\end{equation} 
is the unit future directed timelike normal to the foliation.
The evolution proceeds along the streamlines of the vector
field $\partial_t = \alpha n^\mu \partial_\mu +\beta^\mu \partial_\mu$
determined by the lapse $\alpha$ and shift $\beta^\mu =(0,\beta^i)$.

In this formulation the constraints are
\begin{eqnarray}
\label{eq:GaugeConstraint}
  {\cal C}^\mu := \Gamma^\mu - F^\mu
\label{eq:hc}
\end{eqnarray}
where $\Gamma^\mu$ is related to the Christoffel symbols
by $\Gamma^\mu =g^{\rho\sigma} \Gamma^\mu_{\rho\sigma}$
and $F^\mu(g,x)$ are harmonic gauge source functions~\cite{fried}.

When the harmonic constraints are satisfied, i.e. ${\cal C}^\mu=0$, the Einstein equations 
reduce to a system of quasilinear wave equations for the spatial metric $h_{ij}$,
shift $\beta^i$ and  lapse $\alpha$ which take the form~\cite{friedren}
\begin{eqnarray}
\label{eq:ADMHarm2metric}
- g^{\mu\nu} \partial_\mu \partial_\nu h_{ij} &=& S_{ij}
\\
\label{eq:ADMHarm2shift}
\frac{1}{\alpha^2} \left( \partial_t - \beta^j \partial_j \right)^2 \beta^i
- h^{jk} \partial_j \partial_k \beta^i
              &=& S^i
\\
\label{eq:ADMHarm2lapse}
\frac{1}{\alpha^2} \left( \partial_t - \beta^j \partial_j \right)^2
      \alpha - D_j D^j \alpha 
              &=&  S,
\end{eqnarray}
where $D_i$ is the covariant derivative associated with $h_{ij}$
and the right-hand-side $S$-terms do not enter the principal part.
In detail, these terms are
\begin{eqnarray}
\label{eq:ADMHarm2metricSource}
S_{ij} &=&
\frac{2}{\alpha^2} K_{ij} \left(\partial_t \alpha - \Lie{\beta} \alpha \right)
+\frac{2}{\alpha^2} D_i \alpha D_j \alpha
\\ \nonumber &&
- 2 D_{(i} \left[ h_{j)k} h_\nu^k F^\nu \right]
+ \frac{4}{\alpha^3} D_{(i} \alpha h_{j)k} \left( \partial_t \beta^k - \beta^\ell \partial_\ell \beta^k 
\right)
\\ \nonumber &&
+\frac{4}{\alpha^2} \left( \partial_{(j} \beta^k \right) \partial_t h_{i)k}
+ \frac{2}{\alpha^2} h_{\ell(i} \left( \partial_{j)} \beta^k \right)
\left(\partial_k \beta^\ell \right)
\\ \nonumber &&
- \frac{2}{\alpha^2} \left( \partial_{(j} \beta^k \right) 
 \Lie{\beta} h_{i)k} - \frac{2}{\alpha^2} \left( \partial_{(j} \beta^k \right)
\left( \partial_\ell h_{i)k} \right)
\\ \nonumber &&
+ 4 K_{ik} K^k_j - 2 K_{ij} K - 2 \gamma^\ell_{ki} h_{n\ell} g^{km} \gamma^n_{mj}
- 4 \gamma^\ell_{km} h^{kn} h_{\ell(i} \gamma^m_{j)n} + B_{ij}
\\
\nonumber && \\
\label{eq:ADMHarm2shiftSource}
S^i &=& 4 \left( K^{i\ell} - K h^{i\ell} \right) D_\ell \alpha
\\ \nonumber &&
-2 \alpha \left( K^{m\ell} - K h^{m\ell} \right) \gamma^i_{\ell m}
- \left( \partial_\ell \beta^i \right) \gamma^\ell
\\ \nonumber &&
+ \left( \partial_\ell \beta^i \right) D^\ell \log \alpha
\\ \nonumber &&
+ 2 \alpha n_\nu F^\nu \left[ \gamma^i - D^i \log\alpha - h^i_\nu F^\nu \right]
\\ \nonumber &&
- 2 \alpha K h^i_\nu F^\nu - D^i \left( \alpha n_\nu F^\nu \right)
-\left( \partial_t - \beta^k \partial_k \right) \left( h^i_\nu F^\nu \right) + B^i
\\
\label{eq:ADMHarm2lapseSource}
S &=& - \alpha \bigg[ K_{ij} K^{ij}
 - 4 K n_\nu F^\nu - 2 \left( n_\nu F^\nu \right)^2
-2 K^2 \bigg]
\\ \nonumber &&
+ \left(\partial_t - \beta^k \partial_k \right) \left( n_\nu F^\nu \right) + B ,
\end{eqnarray}
where $ \gamma^k_{ij}$ are the Christoffel symbols associated with $D_i$,
$ \gamma^k= h^{ij} \gamma^k_{ij}$, $K_{ij} = (1/2\alpha)(\partial_t-\Lie{\beta})h_{ij}$
is the extrinsic curvature of the Cauchy foliation and the
quantities $B, B^i, B_{ij}$ are constraint modification terms defined in
(\ref{eq:ConstraintDampingB})-(\ref{eq:ConstraintDampingBij})
which are used to stabilize the constraint propagation system.
With the addition of constraint modification,
(\ref{eq:ADMHarm2metric})-(\ref{eq:ADMHarm2lapse})
correspond to equations (2.33), (2.36) and (2.38)
of~\cite{friedren}, as corrected for typographical errors. See~\cite{pthesis}
for further details.

Constraint preservation for the Cauchy problem follows from the system
of homogeneous wave equations satisfied by the harmonic
constraints (\ref{eq:GaugeConstraint}). Constraint
propagation can be extended to the initial-boundary value problem by
implementing the hierarchy of Sommerfeld outer boundary conditions presented
in~\cite{bkw,isol}.
However, this has not yet been incorporated in the version
of the code being used here and we maintain constraint preservation by causally isolating
the region of interest from the outer boundary.

For the purpose of using the method of lines to apply a Runge-Kutta time
integrator, we re-write the system
(\ref{eq:ADMHarm2metric})-(\ref{eq:ADMHarm2lapse}) as
first differential order in time and second differential order in space  
by using the timelike vector field
\begin{eqnarray}
\label{eq:ADMHarmDefN}
 {\tilde n}^\nu &=& \frac{1}{\alpha} \left( \delta^\nu_0 - w \beta^\nu \right)
\end{eqnarray}
to introduce the auxiliary variables
\begin{eqnarray}
\label{eq:ADMHarmDefAlapse}
 \Pi &:=& \Lie{\tilde n}\alpha = {\tilde n}^\mu \partial_\mu \alpha
\\
\label{eq:ADMHarmDefAshift}
\Pi^i &:=& \Lie{\tilde n}  \beta^i = 
{\tilde n}^\mu \partial_\mu \beta^i - \beta^k \partial_k {\tilde n}^i
\\
\label{eq:ADMHarmDefAmetric}
\Pi_{ij} &:=& \frac{1}{2} \Lie{\tilde n} g_{ij} = 
\frac{1}{2} \left[ 
{\tilde n}^\mu \partial_\mu h_{ij} + g_{i \mu} \partial_j {\tilde n}^\mu
+ g_{j \mu} \partial_i {\tilde n}^\mu
\right].
\end{eqnarray}
Here the function $w$ is chosen to be unity everywhere except near the
boundary, where it smoothly goes to zero.  
Insertion of (\ref{eq:ADMHarmDefN}) into 
(\ref{eq:ADMHarmDefAlapse})-(\ref{eq:ADMHarmDefAmetric}) 
gives the evolution equations for the lapse, shift and 3-metric,
\begin{eqnarray}
\label{eq:ADMHarm1lapse}
\partial_t  \alpha &=&  \alpha \Pi
+ w \beta^i \partial_i  \alpha 
\\
\partial_t  \beta^i &=& \alpha \Pi^i + \frac{w}{\alpha} \beta^i\beta^k \partial_k \alpha 
       -  \beta^i\beta^k \partial_k w
\label{eq:ADMHarm1shift}
\\
\partial_t h_{ij} &=& 2 \alpha \Pi_{ij} 
   -  \alpha \left[ {{\tilde n}}^k \partial_k h_{ij}
                  + g_{i\mu} \partial_j {\tilde n}^\mu
                  + g_{\mu j} \partial_i {\tilde n}^\mu \right] .
\label{eq:ADMHarm1metric}
\end{eqnarray}
The evolution equations for the auxiliary $\Pi$-variables then follow from
the first time derivatives of
(\ref{eq:ADMHarm1lapse})-(\ref{eq:ADMHarm1metric}), after using 
(\ref{eq:ADMHarm1lapse})-(\ref{eq:ADMHarm1metric})
and (\ref{eq:ADMHarm2metric})-(\ref{eq:ADMHarm2lapse}) 
to eliminate first and second time-derivatives of the 
metric variables.

We modify the evolution system by terms vanishing modulo the
constraints based upon the results presented in~\cite{bsw}.
For this purpose we set
\begin{eqnarray}
\label{eq:ConstraintDampingOrig}
A^{\mu\nu} &=& - \frac{a_1}{\sqrt{-g}} 
     {\cal C}^\alpha \partial_\alpha  \left(\sqrt{-g} g^{\mu\nu}\right)
+ \frac{a_2  {\cal C}^\alpha \nabla_\alpha t}{\varepsilon 
+ e_{\rho\sigma} {\cal C}^\rho {\cal C}^\sigma} {\cal C}^\mu {\cal C}^\nu
- \frac{a_3} { \sqrt{-g^{tt}}} {\cal C}^{(\mu} \nabla^{\nu)} t
\end{eqnarray}
where $\varepsilon$ is a small positive number, $a_i$ are
positive parameters and
\begin{equation}
e_{\mu\nu} = n_\mu n_\nu + h_{\mu\nu}
\label{eq:reimmet}
\end{equation}
is a Riemannien 4-metric. In terms of
\begin{equation}
\tilde A_{\mu\nu} = g_{\mu\sigma} g_{\nu\rho} A^{\sigma\rho}
- \frac{1}{2} g_{\mu\nu} g_{\rho\sigma} A^{\rho\sigma}
\end{equation}
the $B$-terms added to
(\ref{eq:ADMHarm2metricSource})-(\ref{eq:ADMHarm2lapseSource})
for  constraint modification are
\begin{eqnarray}
\label{eq:ConstraintDampingB}
B &=&
   \frac{1}{\alpha} \left( - \beta^i\beta^j \tilde A_{ij} 
+ 2 g_{ij} \beta^j \beta^k g^{il} \tilde A_{kl}
+ 2 g_{ij} \beta^j g^{ik} \tilde A_{tk} \right)
\\
\label{eq:ConstraintDampingBi}
B^i &=& -2  \left( \beta^j g^{ik} \tilde A_{kj} + g^{ij} \tilde A_{tj} \right)
\\
\label{eq:ConstraintDampingBij}
B_{ij} &=& -2 \tilde A_{ij} .
\end{eqnarray}

\subsection{Numerical implementation}

The evolution algorithm uses fourth order centered finite differencing
and sixth order Kreiss-Oiliger type dissipation on a cubic Cartesian grid.
The outer boundary points are updated using a
Sommerfeld algorithm provided by the Cactus toolkit~\cite{cactus_web}, while
we use summation-by-parts finite difference operators to update the neighboring
three points along the normal Cartesian axis.

In addition, following the lead of the BSSN method for treating punctures, 
we compute the spatial derivatives of the 3-metric $h_{ij}$ 
using the conformal rescaling
\begin{eqnarray}
    \tilde h_{ij} = h^{-1/3} h_{ij} 
\end{eqnarray}
to obtain
\begin{eqnarray}
\partial_k h_{ij} &=& \tilde h_{ij} \partial_k h^{1/3} +  h^{1/3} \partial_k \tilde h_{ij}
\\
\partial_k \partial_\ell h_{ij} &=& \tilde h_{ij} \partial_k \partial_\ell h^{1/3}
+ \left( \partial_\ell  h^{1/3} \right) \left( \partial_k \tilde h_{ij} \right)
\\ \nonumber &&
+ \left( \partial_k h^{1/3} \right)
  \left( \partial_\ell \tilde h_{ij} \right)
+ h^{1/3} \partial_k \partial_\ell \tilde h_{ij},
\end{eqnarray}
where derivatives of $\tilde h_{ij}$
are computed using finite difference operators.
The derivatives of the conformal factor $h^{1/3}$ are computed using
\begin{eqnarray}
\partial_k h^{1/3} &=& h^{1/3} \partial_k \left[\log \left(h^{1/3}\right)\right] \\
\partial_k \partial_\ell h^{1/3} &=&
 h^{1/3}
 \partial_k
\left[\log \left(h^{1/3}\right)\right]
 \partial_\ell
\left[\log \left(h^{1/3}\right)\right]
+ h^{1/3} \partial_k \partial_\ell 
\left[\log \left(h^{1/3}\right)\right] ,
\end{eqnarray}
where the finite difference operators are applied to the quantity
$
\left[\log \left(h^{1/3}\right)\right]
$.

In addition to the Cactus infratructure we use
Carpet mesh refinement~\cite{carpet1,carpet2}.
The MOTSs are tracked using the horizon finder developed
by J. Thornburg~\cite{thorn1,thorn2}, which decomposes a
topologically spherical surface into 6 abutting cubical patches.
This finder is based upon
a search algorithm which assumes a star-shaped domain
inside the horizon.

As this is the first example of harmonic evolution of black holes
using the puncture method, as opposed to excision, we adopt a
conservative approach to insure that the puncture never lies
exactly on a grid point. This is made possible in the simulation
of  a head-on collision by offsetting the puncture a half
grid-step from the axis along which the collision
proceeds. In addition, for simulation of a head-on collision,
harmonic gauge forcing is not necessary and we set $F^\mu=0$. 
Further studies and code development would be
necessary to better understand the harmonic application of the
puncture method so that the code could handle generic binary inspirals.
With the present techniques, however, 
we can evolve the head-on collision for time scales
well past the formation of a common apparent horizon and are able
to track the
subsequent penetration of the individual MOTSs. 

\section{Theory of the uniqueness and stability of MOTSs}
\label {sec:theory}

Again consider a 3+1 foliation $x^\alpha=(t,x^i)$ with metric
decomposition 
\begin{equation} 
       g_{\mu\nu}= -n_\mu n_\nu +h_{\mu\nu} ,
\end{equation} 
where $n_\mu$ is the unit future directed timelike normal to the
foliation. In a slice ${\cal M}_\tau := \{t=\tau\}$ consider a 2-surface
${\cal S}$ which is the boundary of a set $\Omega$. We define its
outward unit spacelike normal $N_i$ to point out of $\Omega$. If
${\cal S}$ is defined as the level set of a function $s$, then
\begin{equation}
  \label{eq:2}
  N_i = \frac{1}{\sqrt{D}} \partial_i s
\end{equation}
with
\begin{equation}
  \label{eq:3}
  D = h^{kl}(\partial_k s)(\partial_l s).
\end{equation}
Setting $N_\mu=h_\mu^i N_i$ (so that $N^\mu \nabla_\mu t =0$), this
leads to the further decomposition
 \begin{equation} 
       h_{\mu\nu}= N_\mu N_\nu +s_{\mu\nu}. 
 \end{equation} 
The outgoing null direction normal to
${\cal S}$ is 
\begin{equation} 
               \ell^\mu = n^\mu + N^\mu,
\end{equation} 
where the normalization is determined by the Cauchy slicing. The
expansion in this outgoing null direction is  
\begin{equation} 
             \theta_+= P+H \label{eq:thetap}
\end{equation}    
where 
\begin{equation} 
             H= s^{\mu\nu}\nabla_\mu N_\nu
\end{equation}
is the mean curvature of ${\cal S}$ in the Cauchy slicing,
\begin{equation}    
           P = s^{\mu\nu}\nabla_\mu n_\nu
\end{equation} 
is the 2-trace of the
extrinsic curvature of the Cauchy slicing and $\nabla$ is the covariant
derivative with respect to $g$.

${\cal S}$ is a MOTS  if $\theta_+=0$, i.e. $P+H=0$. Similarly, 
\begin{equation}
                        \theta_-= P-H 
\end{equation}  
is the expansion in the ingoing null direction 
\begin{equation} 
                  k^\mu = n^\mu - N^\mu 
\end{equation} 
normal to ${\cal S}$.  The MOTS ${\cal S}$ is inner trapped if $\theta_- <0$. Since
$\theta_-=\theta_+ -2H =-2H$ for a MOTS, the trapping condition is equivalent to
$H>0$. 

\subsection{Stable and outermost MOTSs}
\label{sec:stable-outerm-mots}

For a given slice ${\cal M}_\tau$ and a MOTS ${\cal  S}\subset {\cal M}_\tau$
one can consider the normal graphs ${\cal S}_u$
of a function $u\in C^\infty({\cal S})$, i.e. the surface
parametrized by
\begin{equation}
  \label{eq:6}
  F_u: {\cal S}\to {\cal M}_\tau : p \mapsto \exp(uN^i)
\end{equation}
where $\exp$ denotes the exponential map of ${\cal M}_\tau$. The
operator $\Theta_+ : C^{\infty}({\cal S}) \to C^\infty({\cal S})$ that
maps a function $u$ to the pull-back of the value of $\theta_+$ on
${\cal S}_u$ to ${\cal S}$ has linearization given by
\begin{equation}
  L f
  =
  -\Delta f
  - 2 s^{AB} S_A \partial_B f
  + f( \tfrac12 R_{\cal S} - S_AS^A  + D^AS_A - \tfrac12
  \chi^{AB}\chi_{AB} - G_{\mu\nu} \ell^\mu k^\nu)
  \label{eq:linearizatiion}
\end{equation}
with $\Delta$ the surface Laplacian of ${\cal S}$, $S_\mu = N^\nu \nabla_\mu
n_\nu $, $R_{\cal S}$ denotes the scalar curvature of ${\cal S}$,
$\chi_{\mu\nu} = \nabla_\mu \ell_\nu$, $G_{\mu\nu}$ denotes the
Einstein tensor of the space-time and capital letters refer to
intrinsic coordinates $x^A$ for ${\cal S}$. In particular $s^{AB}$ denotes
the inverse of the tangential projection of $s_{\mu\nu}$ to ${\cal S}$.

Although $L$ is not self-adjoint, there exists a real eigenvalue
$\lambda({\cal S})\in \sigma(L)$, which is the unique minimizer for
the real part in the spectrum $\sigma(L)$ of $L$. This eigenvalue is simple, and the
corresponding eigenfunction $\phi$ can be chosen with a definite
sign. If $\lambda({\cal S})\geq 0$ we say that ${\cal S}$ is
\emph{stable}, if $\lambda({\cal S})>0$ it is called \emph{strictly
  stable}.  We refer to
\cite{Andersson-Mars-Simon:2005,Andersson-Mars-Simon:2008} for further
details.

For the following, we require the surfaces ${\cal S}$ in question to
be either bounding, ${\cal S} = \partial \Omega$, or bounding with respect
to an interior boundary $\partial {\cal M}$, that is ${\cal S}
= \partial\Omega\setminus \partial {\cal M}$. In both cases, we write
${\cal S} = \partial^+\Omega$ or refer to ${\cal S}$ being an outer
boundary in this situation. In the scenario considered here, the inner
boundary exists and is formed by trapped surfaces enclosing the punctures.

A MOTS ${\cal S} = \partial^+ \Omega$ is called
\emph{outermost} if for all other MOTSs ${\cal S}' = \partial^+ \Omega'$
with $\Omega'\supset\Omega$ it follows that $\Omega'=\Omega$. In other
words, there are no MOTSs on the outside of ${\cal S}$. 

In \cite{Andersson-Metzger:2009} it was shown that if ${\cal M}_\tau$
contains bounding outer trapped surfaces, as is the case if
${\cal M}_\tau$ is asymptotically flat, then there exists an outermost MOTS
${\cal S}^\text{out}$ that bounds the trapped region in
${\cal M}_\tau$. This means, it is the enclosure of the region that
contains outer trapped surfaces. Note that ${\cal S}^\text{out}$ is
not necessarily connected. All components of ${\cal S}^\text{out}$ are
stable, and ${\cal S}^\text{out}$ has area bounded uniformly from above by
a constant depending only on the geometry of the slice ${\cal
  M}_\tau$. Furthermore, it has the property that there exists a
positive $\delta>0$, again depending only on the geometry of the slice
${\cal M}_\tau$, such that any geodesic starting on ${\cal
  S}^\text{out}$ in direction of its outward normal $N^i$ does not
intersect ${\cal S}^\text{out}$ within distance $\delta$. In
particular, two distinct components of ${\cal S}^\text{out}$ have
distance at least $\delta$. The constants mentioned depend in
particular on an intrinsic curvature bound and on bounds for the
second fundamental form $\nabla_\mu n_\nu$ and its derivatives. For
the details of these estimates and all the dependencies of the
constants we refer to \cite{Andersson-Metzger:2009}. Similar results
have also been derived in \cite{Eichmair:2007,Eichmair:2008}. In the situation
considered here, Galloway \cite{Galloway:2008} established that ${\cal
  S}^\text{out}$ is a union of topological spheres. In
\cite{Andersson-Metzger:2005} it has been furthermore established that
the second fundamental form $\nabla_iN_j$ of a stable MOTS ${\cal S}$
is bounded in the supremum norm, provided the geometry of the slice is
bounded.

\subsection{The maximum principle for MOTSs}
\label{sec:maxim-princ-mots}

A useful tool in the analysis of MOTSs is the strong maximum
principle. To state it, assume that ${\cal S}_\alpha$ for $\alpha=1,2$ are two
connected $C^2$-surfaces with outer normals $N^i_\alpha$. Assume further
that there is a point $p$ such that ${\cal S}_1$ and ${\cal S}_2$
touch at $p$. If the outer normals agree, $N^i_1=N^i_2$, at $p$ and
${\cal S}_2$ lies to the outside of ${\cal S}_1$, that is in the direction
of $N^i_1$, and furthermore
\begin{equation}
  \sup_{{\cal S}_1} \theta_+[{\cal S}_1] \leq \inf_{{\cal S}_2}\theta_+[{\cal S}_2]
\end{equation}
then ${\cal S}_1={\cal S}_2$.
This version can be found in \cite[Proposition 2.4]{Andersson-Metzger:2009} or in
\cite{Ashthekar-Galloway:2005}. It implies in particular that two distinct MOTSs
${\cal S}_1$ and ${\cal S}_2$ can not touch in such a way that their
normals point in the same direction and one is enclosing the other.

The strong maximum principle provides an interpretation of strict
stability. Assume that ${\cal S}$ is a strictly stable MOTS with
outward normal $N^i$. Let $\phi$ be the principal eigenfunction of
$L$. Deforming ${\cal S}$ in direction of the vector field $\phi N^i$
then yields a foliation of a tubular neighborhood $U$ of ${\cal S}$
with the following properties. First $U\setminus {\cal S}=U^- \cup
U^+$ where $N^i$ points into $U^+$. Moreover, $U^+$ is foliated by
surfaces with $\theta_+>0$ and $U^-$ is foliated by surfaces with
$\theta_+<0$. The maximum principle then implies that there are no
surfaces ${\cal S}'\subset U^+$ which bound relative to ${\cal S}$ and
have $\theta_+[{\cal S}']\leq 0$. Furthermore, there is no MOTS in $U$
with outward normal aligned with that of ${\cal S}$.

\subsection{Evolution of MOTSs to MOTTs}
\label{sec:evolution-mots-mtts}

Regarding the evolution of MOTSs there are different approaches. In
\cite{Andersson-Mars-Simon:2005} it was shown that a strictly stable
MOTS ${\cal S}$ can locally be continued to a smooth space-time track
of MOTSs, i.e. a marginally outer trapped tube. More precisely,
for a given $\bar \tau$ such that ${\cal M}_{\bar\tau}$ contains a strictly
stable MOTS ${\cal S}_{\bar\tau}$ there exists $\varepsilon>0$ such that
for all $\tau\in (\bar \tau-\varepsilon,\bar \tau+\varepsilon)$ there is a
stable MOTS ${\cal S}_\tau$ in ${\cal M}_\tau$ such that the ${\cal S}_\tau$
form a smooth space-like manifold. To emphasize the role of the
stability operator in this picture, we recall the argument from
\cite{Andersson-Mars-Simon:2005}. Assume that ${\cal S}_\tau$ is a smooth
family of MOTSs passing through ${\cal S}_{\bar\tau}$. Then we can
parametrize this tube by a map
\begin{equation}
  \label{eq:7}
  F^\mu:(\bar\tau - \varepsilon, \bar\tau + \varepsilon) \times {\cal
    S}_{\bar\tau} 
  \to \bigcup_{\tau\in(\bar\tau - \varepsilon, \bar\tau +
    \varepsilon)} {\cal M}_\tau
\end{equation}
such that $\frac{\partial
  F^\mu}{\partial\tau} = V^\mu$, where $V^\mu$ is perpendicular to
${\cal S}_\tau$ at each point along the tube. Note that $V^\mu$ has the
decomposition
\begin{equation}
  \label{eq:4}
  V^\mu = \alpha n^\mu + \gamma N^\mu
  =
  \alpha (n^\mu + N^\mu) + (\gamma -\alpha) N^\mu
  =
  \alpha \ell^\mu + f N^\mu ,
\end{equation}
where as before $\alpha$ denotes the lapse function of the
slicing. Calculating the change of $\theta_+$ under the deformation by
$V^\mu$ at time $\bar\tau$, we thus obtain
\begin{equation}
  \label{eq:5}
  \delta_{V^\mu} \theta_+[{\cal S}_{\bar\tau}]
  =
  \delta_{\alpha \ell^\mu} \theta_+[{\cal S}_{\bar\tau}]
  + \delta_{fN^\mu} \theta_+[{\cal S}_{\bar\tau}]  
  =
  - \alpha W  + Lf ,
\end{equation}
where the first contribution is calculated via the Raychaudhuri
equation with
\begin{equation}
  \label{eq:8}
  W = \chi^{AB}\chi_{AB} + G_{\mu\nu}\ell^\mu\ell^\nu,
\end{equation}
and the second part is just the definition of the stability
operator (\ref{eq:linearizatiion}).
Since $V^\mu$ is tangent to a MOTT we have $\delta_{V^\mu}
\theta_+[{\cal S}_{\bar\tau}]=0$ and thus
\begin{equation}
  \label{eq:1}
  Lf = \alpha W.
\end{equation}
The operator $L$ and the function $W$ are given by the
geometry of ${\cal S}_{\bar\tau}$ and the space-time geometry of ${\cal
  M}_{\bar\tau}$, whereas $f$ is a function determined by
(\ref{eq:5}). If ${\cal S}_{\bar\tau}$ is strictly stable then
$L$ is invertible. Thus the previous calculation can be turned around
to conclude the existence of the desired MOTT. The causal structure of
the tube follows by the observation that the null energy condition
implies $W\geq 0$. From stability one then obtains $f\geq 0$ via 
equation~(\ref{eq:1}) as in \cite[Lemma 3]{Andersson-Mars-Simon:2005}.

A similar argument was used in \cite{AMMS:2009} to construct a MOTT
through ${\cal S}_{\bar\tau}$ in the case it is stable but not strictly
stable assuming $W\geq 0$ and $W\not\equiv 0$. 

A different approach is to take the outermost MOTS ${\cal
  S}^\mathrm{out}_\tau$ of all slices ${\cal M}_\tau$ and define the
apparent horizon of the slicing as ${\cal H}=\bigcup_\tau {\cal
  S}^\mathrm{out}_\tau$. In generic space-times ${\cal H}$ is smooth up
to a discrete set of outward jumps. See \cite{AMMS:2009} for the
details and the particular notion of genericity used therein. The same
reference also provides the causal character of ${\cal H}$, namely
that ${\cal H}$ is achronal.  Moreover, if $\Omega_\tau$ denotes the
interior of ${\cal S}^\mathrm{out}_\tau$, then $J^+(\Omega_\tau) \cap
{\cal M}_\sigma \subset \Omega_\sigma$ for all $\tau < \sigma$. Here
$J^+(\Omega_\tau)$ denotes the causal future of $\Omega_\tau$. In
particular, if there is a MOTS at an initial time $\bar\tau$ then it
will persist for all times.

\subsection{Approaching MOTSs}
\label{sec:approaching-mots}

Assume that the space-time and the slicing are completely
regular. Then the constant $\delta$, the bound for the area, and
the length of the second fundamental form of ${\cal S}^\mathrm{out}$
remains uniformly controlled. If in this situation two components of a
MOTS are closer than $\delta$, neither of these two components can be
part of ${\cal S}^\mathrm{out}$. In particular, the evolution of ${\cal
  S}^\mathrm{out}$ will be discontinuous at some stage of the
evolution. By the causality of ${\cal H}$, the jump is outward, and a
new outermost MOTS has formed outside the tubes of the two original
ones. Generically, after the jump time the jump target will split into
two branches of MOTTs, a stable branch traveling outward and an
unstable branch traveling inward. For a complete discussion of this
jump in the outermost MOTS, we refer to \cite{AMMS:2009}. An important
fact to point out is that if two MOTSs are close to each other then
there is no reason to expect that they be stable, in contrast to the
stability of ${\cal S}^\mathrm{out}$.

Note that it is not known whether the area of $\mathcal{S}^\text{out}$ is 
monotonic across the jumps. For MOTSs in general, including 
$\mathcal{S}^\text{out}$, it is not even clear whether monotonicity of area 
should be expected along smooth pieces. However, if the MOTS has positive
mean curvature, so that $H=-P >0$, then its expansion in any
outward spacelike direction $N^\mu +\alpha n^\mu$, where $0\le \alpha <1$,
must be positive. As a consequence, the continuous portion of the MOTT traced
out by a stable MOTS with $H>0$ must have a monotonically increasing area.
Such MOTSs will also be inner trapped, i.e. $\theta_- =P-H <0$.

\subsection{Exterior osculation of MOTSs}
\label{sec:exter-oscul-mots}
The following theoretical observations pertain to the collision
between the MOTSs of a large and small black hole as they first
touch. Note that this situation is not prevented by the strong maximum
principle. However, at the time of first contact, a common horizon
already has formed according to the previous
section~\ref{sec:approaching-mots}.

In (\ref{eq:thetap}), the contribution to $\theta_+$ from
$P$ is common to both MOTSs, so that at a common point of osculation we
must have
\begin{equation}
        H_{(small)}=H_{(large)}. 
\label{eq:Hsml2}
\end{equation}
Here the two mean curvatures are defined with respect to the respective outer
normals, which in this case have opposite orientations. Thus the two MOTSs do not
share the same outgoing null direction $\ell$. As a result of (\ref{eq:Hsml2}),
the mean extrinsic curvature radius of the large and small MOTSs must match at
the point of osculation. This can be deceiving in terms of a coordinate picture
of the MOTSs since the connection of the Cauchy slice enters into the mean
curvature. We have
\begin{equation} 
        H=  s^{ij} \partial_i N_j -s^{ij} \Gamma^k_{ij}N_k. 
        \label{eq:hs}
\end{equation}
At the point of osculation $N_{i(small)}=-N_{i(large)}$ so that the second term
has the same magnitude but opposite sign for the small and large MOTSs. Hence
(\ref{eq:Hsml2}) implies  
\begin{equation} 
          s^{ij} \partial_i N_{j(small)}= s^{ij} \partial_i N_{j(large)}
	    -2s^{ij}  \Gamma^k_{ij}N_{k(large)}  
\end{equation}
or 
\begin{equation}
       D^{-1/2}_{(small)} s^{ij} \partial_i \partial_j s_{(small)}=
    D^{-1/2}_{(large)} s^{ij} \partial_i \partial_j s_{(large)} -2s^{ij} 
       \Gamma^k_{ij} N_{k(large)} . 
       \label{eq:shape}
\end{equation}
This relates the
coordinate curvatures $\partial_i \partial_j s$ of the functions describing the
MOTSs, as provided graphically by the code output. More geometrically meaningful
output are  plots of $H$ during the evolution of the large and small MOTSs and plots
of the time dependence of their surface area, at a sequence of times elapsed during
the evolution. An important property is whether $H>0$ so that the MOTSs are
trapped. 

In Thornburgh's apparent horizon finder~\cite{thorn1,thorn2}, 
$s=r-h(y^A)$, where $r$ is a standard radial coordinate
measuring Euclidean distance from some point $x_0^i$ and $y^A$ are spherical
coordinates arising from a six-patch treatment of the unit sphere. Then
\begin{equation}
               \partial_i s = \frac{x^i- x_0^i}{r} -\partial_i h. 
\end{equation} 
In the axisymmetric case corresponding to the head-on collision of black holes, we
must have $\partial_i h=0$ on the symmetry axis and therefore also at the points
where an osculation can occur. Thus, at an an osculation point, we have $
\partial_i s = (x^i- x_0^i)r^{-1}$ and $D=h_{ij}(x^i- x_0^i) (x^j- x_0^j)
r^{-2}=h_{xx}$, where we align the symmetry about the $x$-axis. Thus
$D_{(small)}=D_{(large)}$ at the point of osculation and (\ref{eq:shape}) reduces to
\begin{equation} 
    s^{ij} \partial_i \partial_j s_{(small)}=
        s^{ij} \partial_i \partial_j s_{(large)} 
	-2s^{ij}  \Gamma^k_{ij} \partial_k s_{(large)} ,
\label{eq:shap}
\end{equation}
where $\partial_k s_{(large)}=-\partial_k s_{(small)}$.

\subsection{Interior Osculation}
\label{sec:interior-osculation}

If the MOTS of the small black hole were to remain intact and continue
to completely penetrate the large one then there would again be a
point of osculation between them just as the small one completely passes
inside. At this contact point $p$, the two outer null directions have
the same orientation. Thus $\theta_+ (p)=0$ for the common outgoing
null direction $\ell(p)$. The strong maximum principle implies in this
case that the two MOTSs coincide globally.  Thus the mean
curvatures of the large and small MOTSs must adjust dramatically to
match each other unless full penetration is obstructed by a
singularity or some other feature. One such obstruction of an
artificial nature can arise due to singularity excision. Here we treat
the singularities by punctures, which can restrict either the
simulation or the horizon tracker to the point of half penetration.

According to section~\ref{sec:maxim-princ-mots}, it is impossible in
this case that either of the two osculating MOTSs is strictly stable up to
the time of coincidence. In fact, the normal separation of one MOTS
above the other yields after linearization at the time where they
coincide a function in the kernel of the stability operator. If the
approach is fast enough, this function is non-zero and changes
sign. This implies instability of the two individual MOTSs shortly
before and at the time of coincidence.

\begin{figure}
  \begin{center}
    \includegraphics[width=.5\linewidth]{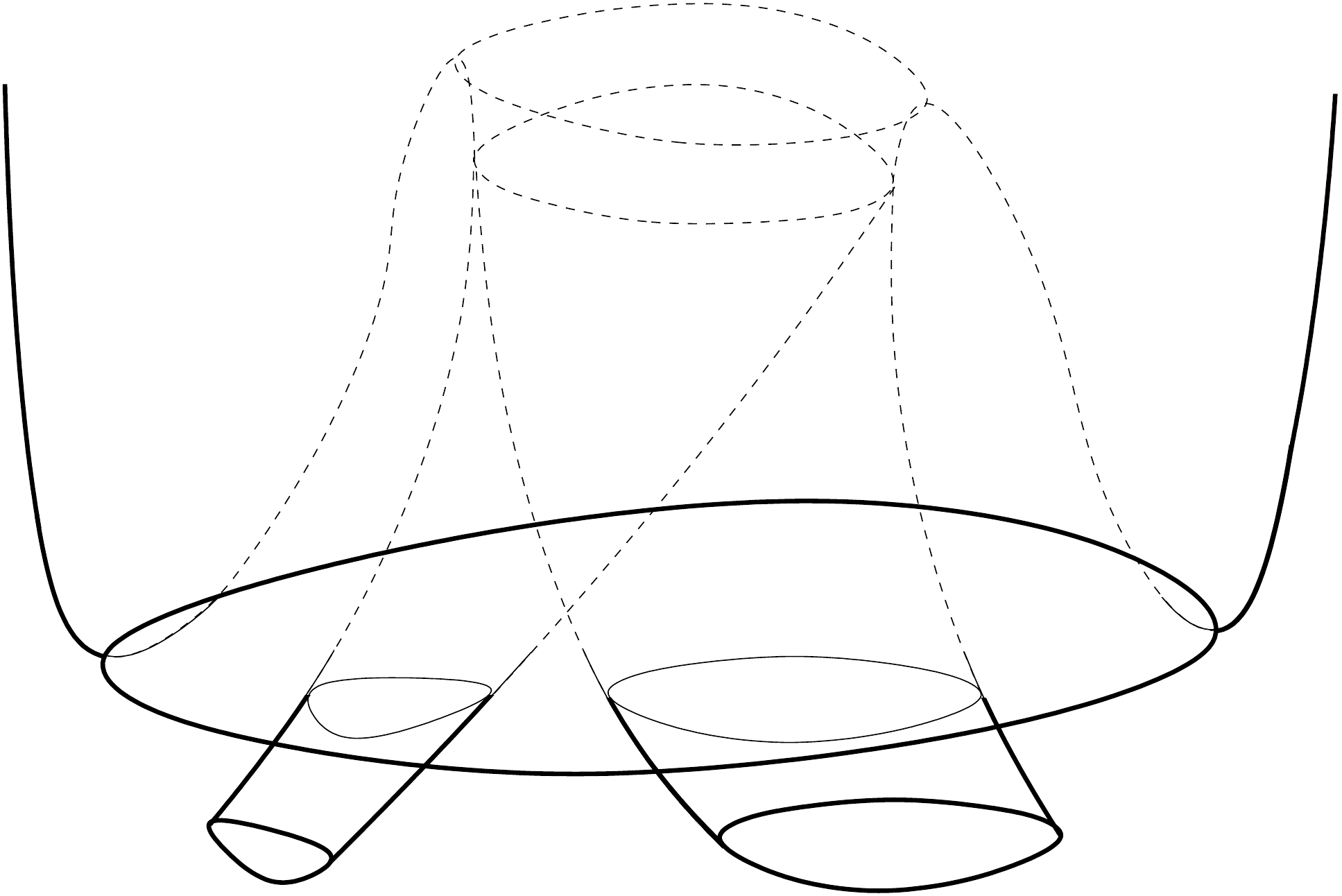}
  \end{center}
  \caption{A possible scenario for the evolution of the trapped
    tubes. The outermost MOTSs are drawn in bold. The common outermost
    MOTS develops before the first time of contact. After the separate MOTTs
    penetrate, the diagram illustrates the speculative scenario that they
    merge and join with the unstable branch from the jump. This speculative
    part of the figure is indicated by the dashed part .}
  \label{fig:situation}
\end{figure}

If the two individual inner MOTSs do coalesce into a single one, this
raises the possibility of a scenario in which the two MOTTs traced out
by the individual MOTSs merge and the resulting MOTT connects to
the unstable branch
of the common horizon, whose stable branch is the MOTT traced out by the apparent
horizon, as described in section~\ref{sec:approaching-mots} and
depicted in Fig.~\ref{fig:situation}. The figure illustrates how the two individual MOTTs
can coalesce and join continuously to the outer apparent horizon.
Although this figure looks
highly suggestive, there is no reason to believe that it is actually
valid. What currently is known is that there always exists an outermost
MOTS, that it can jump, and that at the time of the jump there are two
branches emanating from the jump target. The current state of the
theory is not able to determine how long the unstable MOTSs
continue to exist. See~ \cite{AMMS:2009} for a related discussion.

\section{Simulations}
\label{sec:simulations}

Here we present the results of the simulation
of the head-on collision of two black holes with
initial masses m and m/4. Besides outputting the coordinate shapes
of the MOTSs detected by the horizon finder,
in order to analyze their intrinsic geometrical structure
we also output plots of the
time dependence of
their mean curvature $H$ and total area.
Since this is the first application of the harmonic code described in
Sec.~\ref{sec:compmeth}, we also present convergence tests to establish validity.
Convergence of the constraints
is checked in the region of interest close to the MOTSs. This avoids the expected
loss of convergence near the punctures and near the outer boundary (where
constraint preserving boundary conditions have not been implemented). Because
the run times necessary for the head-on collision are short, errors from the
outer boundary treatment do not effect the region of interest.

\subsection{Initial configuration of the MOTSs}

We use the same time symmetric Brill-Lindquist data to initiate an axisymmetric
head-on-collision
as in~\cite{schkrb}, with the exception that the black holes now have unequal masses.
The conformal flatness of the initial 3-metric provides
a natural choice of Euclidean coordinates $(x,y,z)$.
The initial ``bare'' masses of the punctures, neglecting their interaction, are
$m_{(large)}=4m_{(small)}=0.8M$, corresponding to the Euclidean coordinate radii
of the non-interacting
black holes $r_{(large)}=4r_{(small)}=0.4M$.
We choose the $x$-axis to be the axis of symmetry,
with the punctures corresponding to the two black holes
initially separated by $\Delta x =1.0M$. Due to their interaction at this
separation, the actual masses of the individual black have ratio
$M_{(large)}\approx 3.14M_{(small)}$. (See~\cite{schkrb} for a discussion
of the leading order effect of a finite separation on the physical parameters.)
In order to avoid numerical problems, the punctures  are
offset by a half grid-step from the axis of symmetry.

\begin{figure}
  \begin{center}
    \includegraphics[width=.5\linewidth]{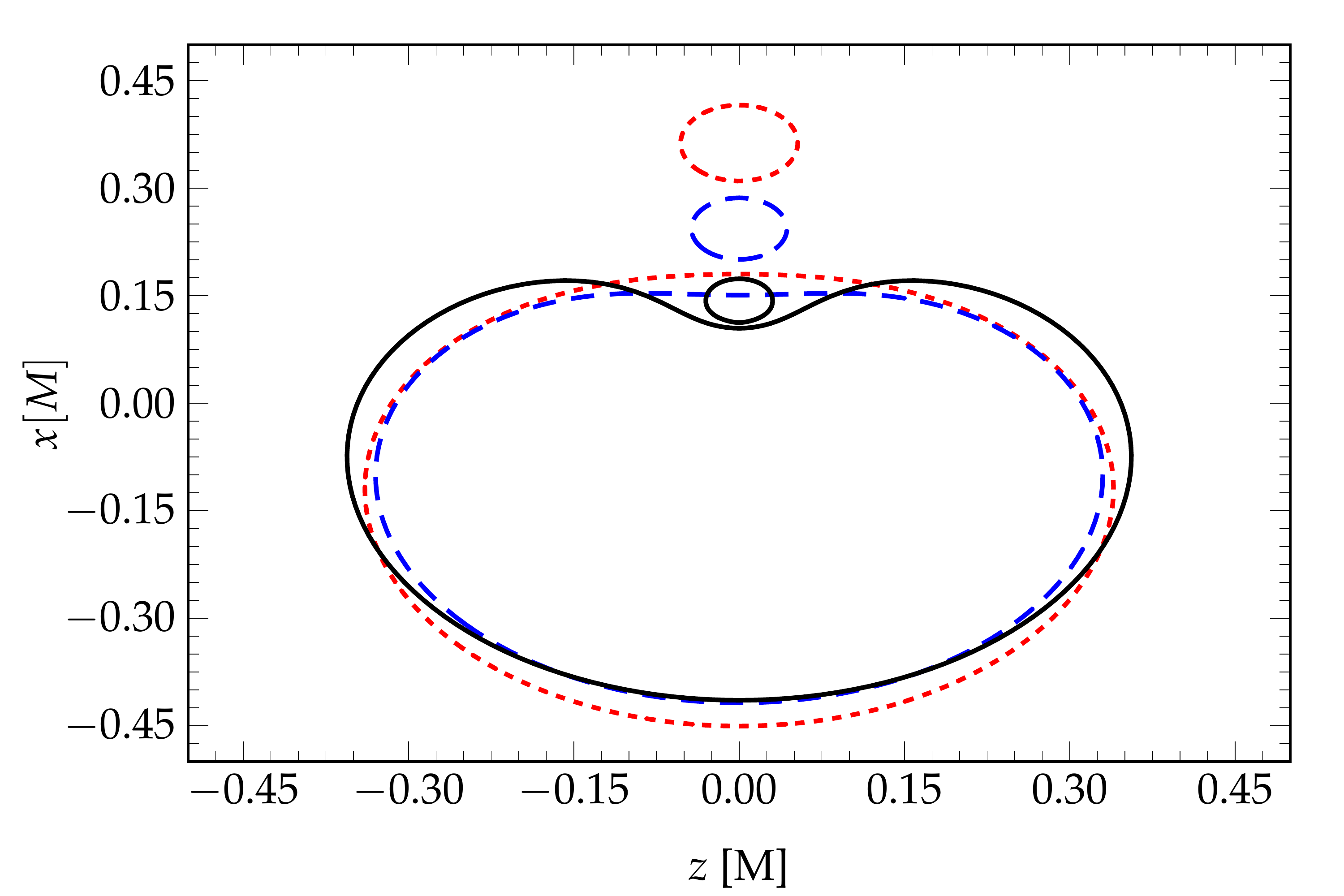}
  \end{center}
  \caption{Cross-sections of the coordinate shapes, in units of $M$, of the individual MOTSs which approach
  each other in a sequence of time-symmetric Cauchy slices. The uniqueness property of minimal surfaces
  keeps the MOTSs from touching and creates considerable coordinate distortion when they are close.}
  \label{fig:example}
\end{figure}

This initial configuration
reduces the initial distortion of the black holes while also reducing the
time required for merger. The sequence of three
time symmetric initial data sets in Fig.~\ref{fig:example}
illustrates the distortion that would arise from smaller separations. It also
illustrates some of the problems which arise
in interpreting the output of the numerical simulations and emphasizes the
importance of monitoring geometric invariants in tracking the evolution of the
MOTSs.

At a moment of time symmetry the MOTS
condition reduces to $H=0$, i.e. from (\ref{eq:hs})
\begin{equation} 
     s^{ij} \partial_i \partial_j s=  
        s^{ij}  \Gamma^k_{ij} \partial_k s . 
\end{equation}
Then for a sequence of initial data for which the two black holes
approach each other,  (\ref{eq:shap}) implies
\begin{equation} 
      s^{ij} \partial_i \partial_j s_{(small)}\rightarrow
         -s^{ij} \partial_i \partial_j s_{(large)}
\end{equation}
at the point of closest approach.
Thus if the coordinate shape of one MOTS appears convex at the point of closest approach then the other
must appear concave with the exact same magnitude of curvature. We see this
effect in the initial data sets shown in Fig~\ref{fig:example}.

Another feature of the sequence of time symmetric data sets is that the MOTSs  cannot touch as
their initial separation is made smaller. This is also a consequence of the
uniqueness theorem for MOTSs. In the time symmetric case both MOTSs satisfy
$\theta_+=\theta_-=0$. If they were to touch then at their common point the outer null
normal to one MOTS would be the same as the inner null normal to the other. But
since both null directions have vanishing expansion, this would violate the
uniqueness theorem. Note that in the time symmetric case, a MOTS is also a
minimal surface so that this result also follows from the uniqueness theorem for
minimal surfaces.

\subsection{Approaching MOTSs}

\begin{figure}
  \begin{center}
    \includegraphics[width=.45\linewidth]{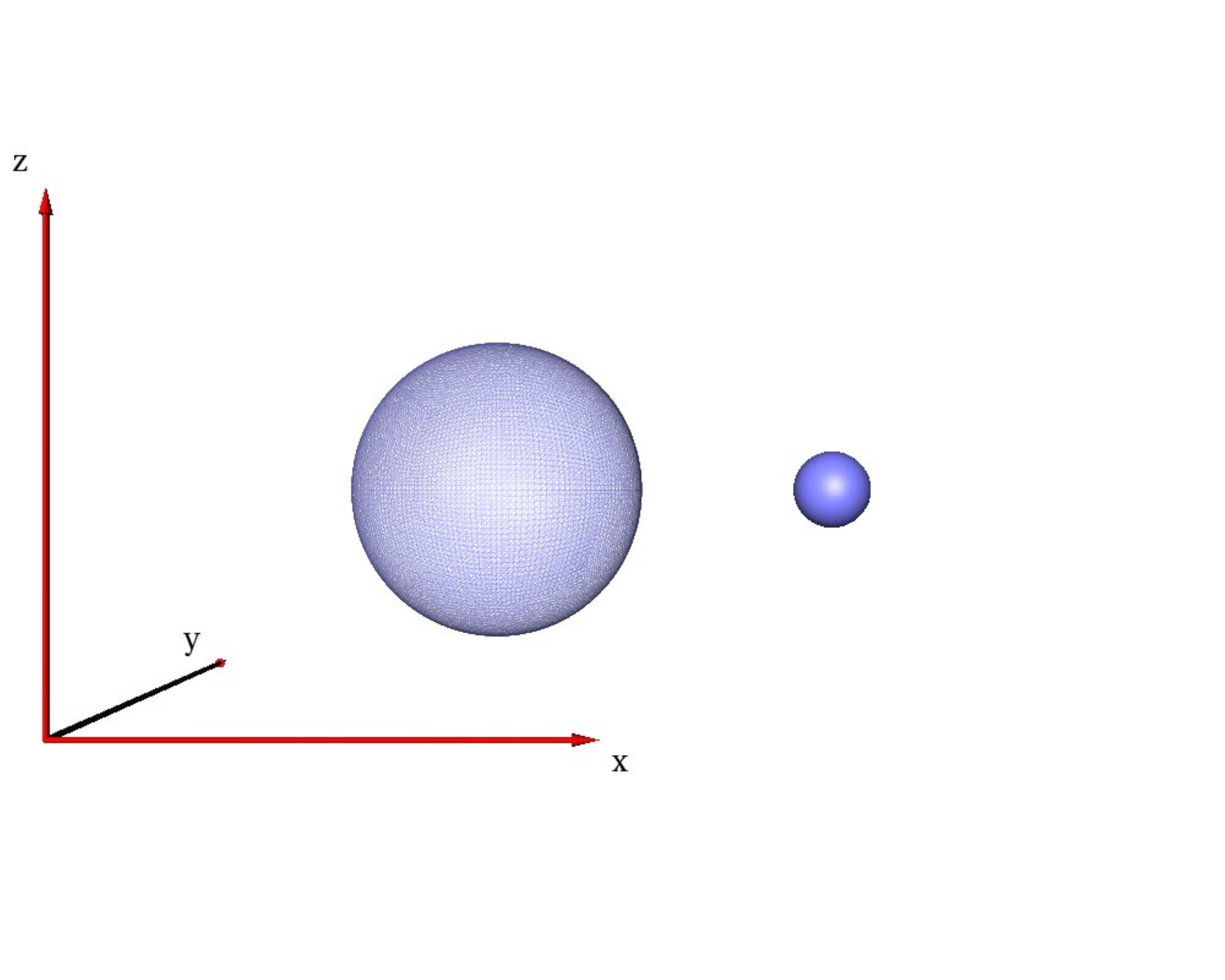}
    \includegraphics[width=.45\linewidth]{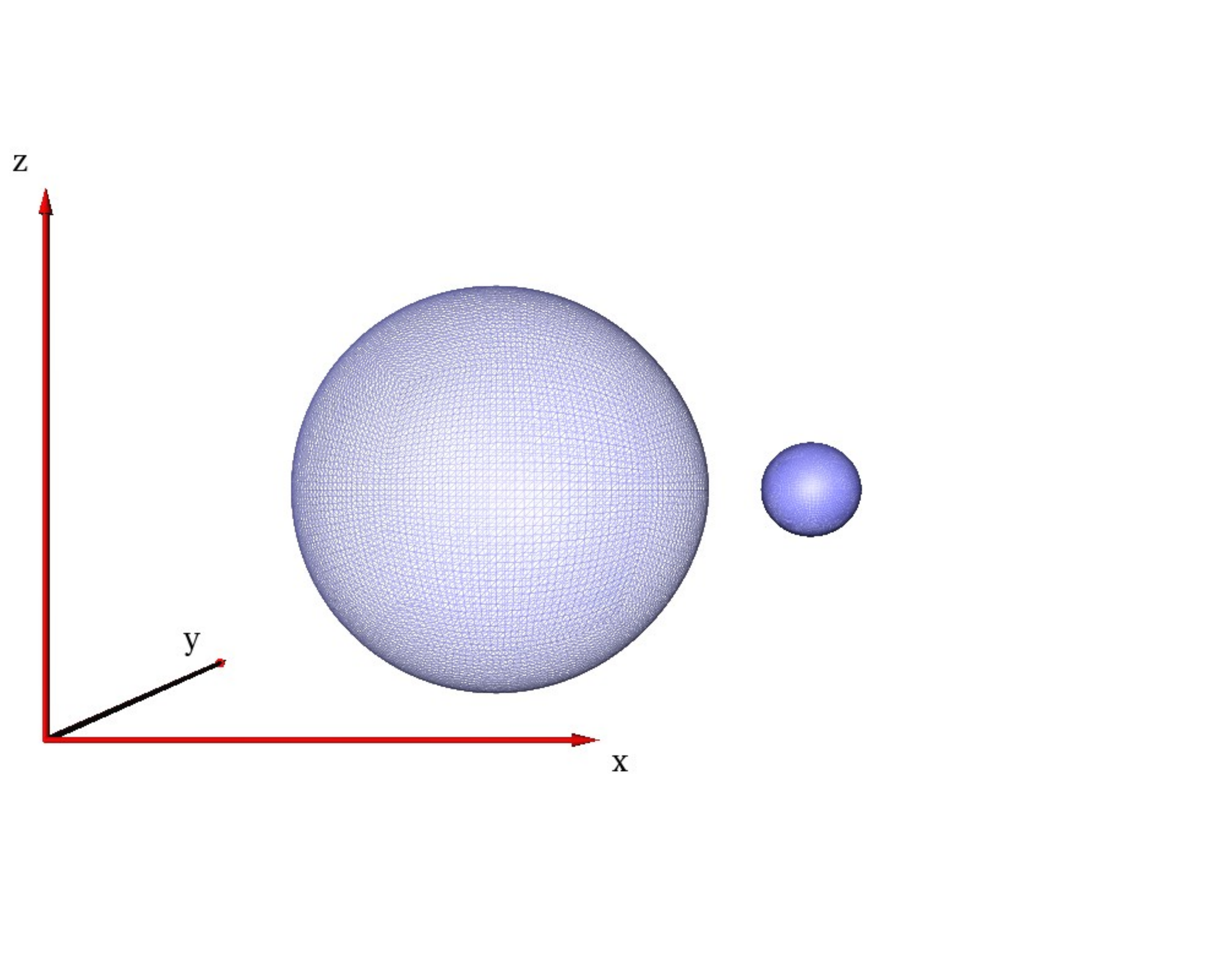}
  \end{center}
  \caption{Coordinate shapes of the individual MOTSs at times
     $t=0.384M$ (left) and $t=1.452M$ (right). At the earlier time, the coordinate shapes
     are almost spherical and then begin to distort as the MOTSs approach.}
  \label{fig:t1}
\end{figure}

As the individual MOTSs approach a common outer horizon (apparent horizon) forms
at $t= 0.144M$, which is before they touch in accord with the theory described in
Sec.~\ref{sec:approaching-mots}. Figure \ref{fig:t1} shows
the individual MOTSs at time $t=0.384M$ and at $t=1.452M$
just after the appearance of
the common horizon. At $t=1.452M$, the bifurcation of the common horizon
has produced a stable branch propagating
outward and an unstable branch propagating inward.
Up to this stage, except for the asymmetry produced
by the unequal black hole masses, 
the results are qualitatively similar to those found in~\cite{schkr}.

\begin{figure}
  \begin{center}
    \includegraphics[width=.45\linewidth]{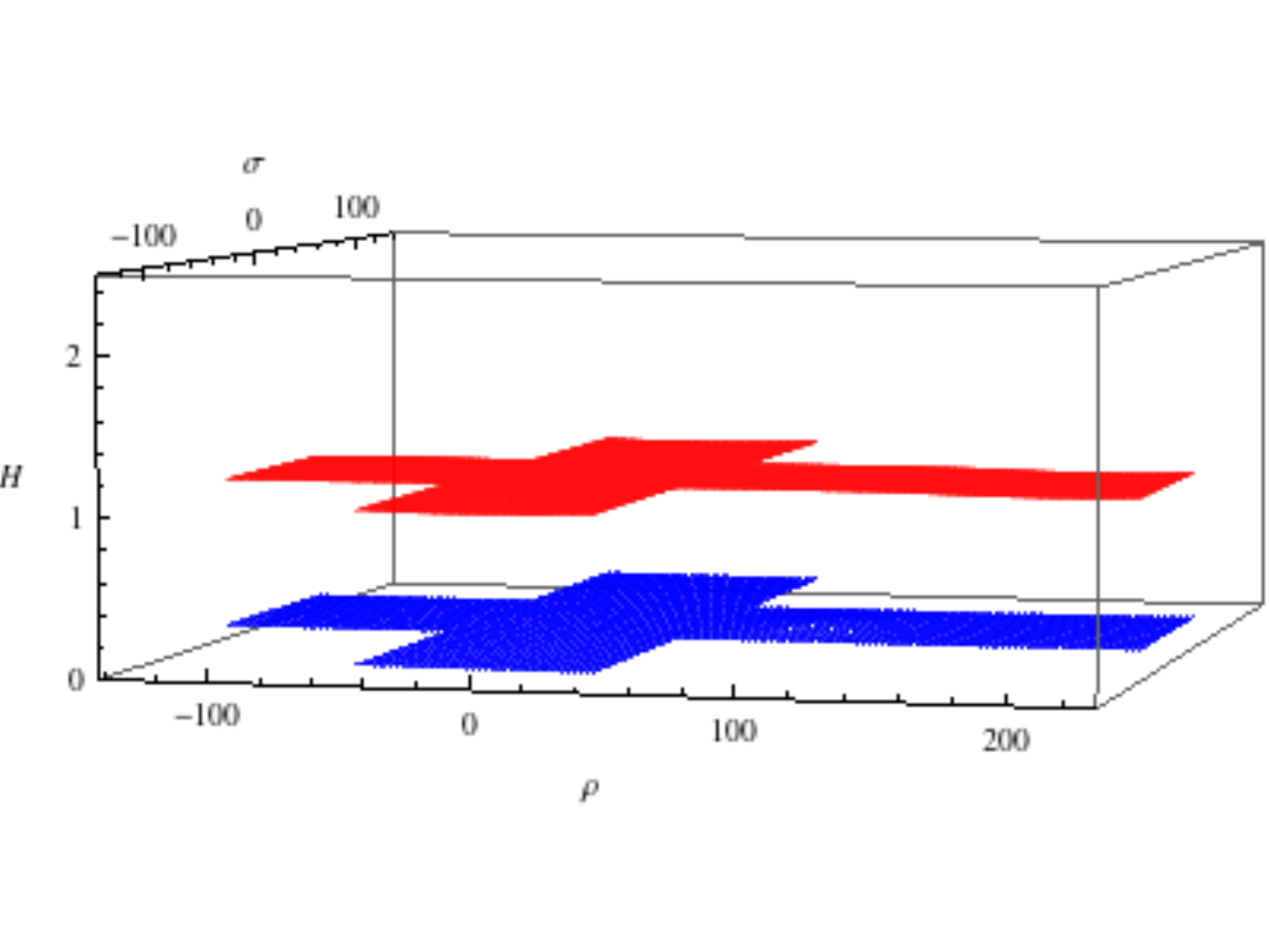}
    \includegraphics[width=.45\linewidth]{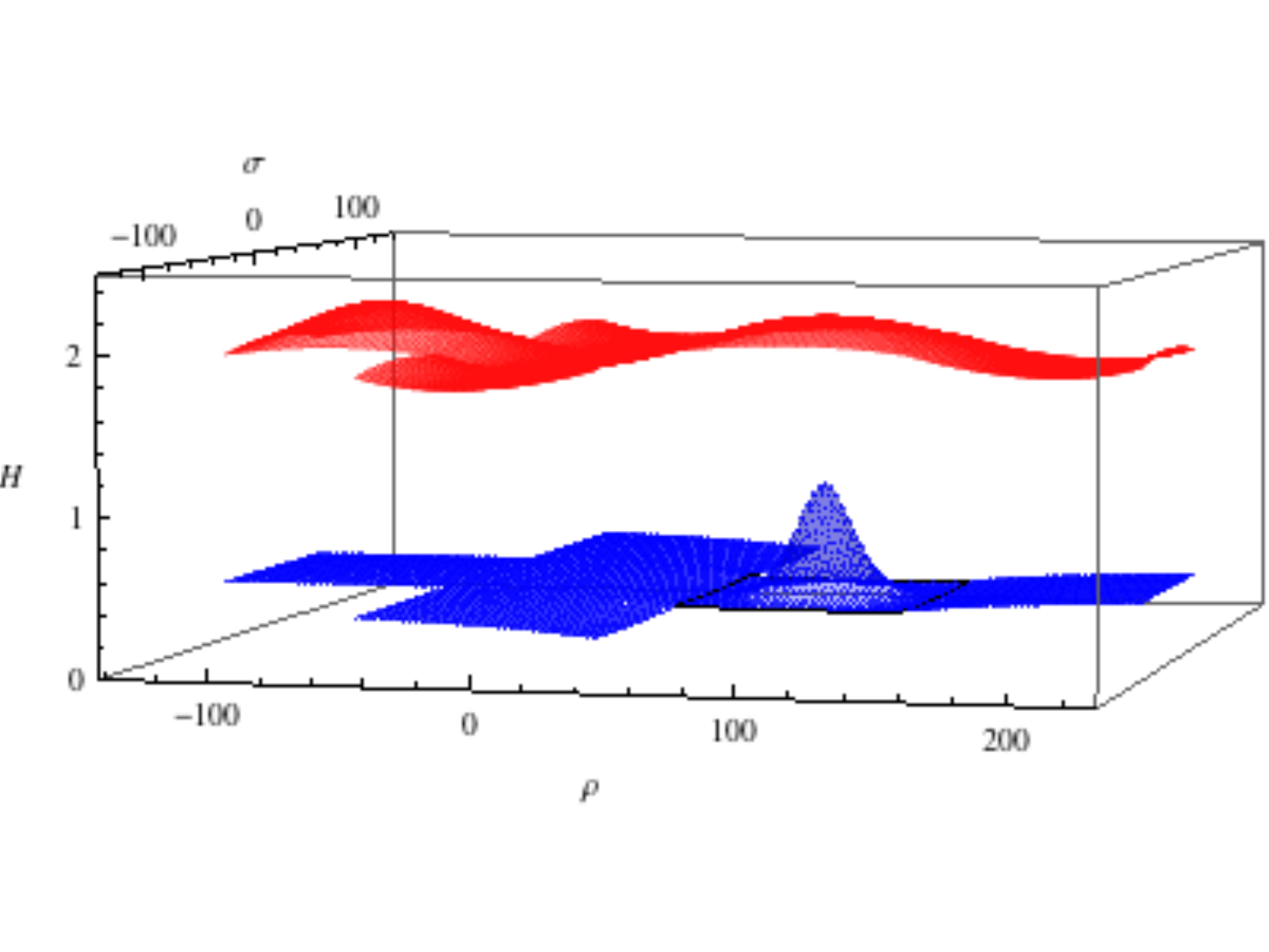}
  \end{center}
  \caption{The mean curvature $H$ of the individual MOTSs at times
          $t=0.384M$ (left) and $t=1.452M$ (right) laid out on the
	 angular coordinates $\rho$ and $\sigma$ of the
          six cubed-sphere patches of the horizon finder. At the earlier time, the
	mean curvatures are fairly constant and have roughly adapted from
	their initial time-symmetric values $H=0$  to the expected
         values $H_{(small)}\approx 4 H_{(large)}$ for the corresponding masses.
         On the right, considerable
        distortion has formed on the front end of the large MOTS. }
  \label{fig:t2}
\end{figure}

Surface plots of the mean curvature $H$ of the individual MOTSs,
on the 6-patch spherical coordinates of the horizon finder,
at times $t=0.384M$ and at $t=1.452M$ are
shown in Figure \ref{fig:t2}. Initially, these mean curvatures are zero,
as a result of the time symmetry. At  $t=0.384M$, they are fairly uniform,
with $H_{(small)}\approx 2$ and  $H_{(large)}\approx 0.5$
in roughly the 4 to 1 ratio of the initial masses.
At $t=1.452M$, the larger MOTS has undergone
significant distortion due to the smaller one, although this is not
evident in its coordinate shape.

\subsection{External osculation}

\begin{figure}[!h]
  \begin{center}
    \includegraphics[width=.45\linewidth]{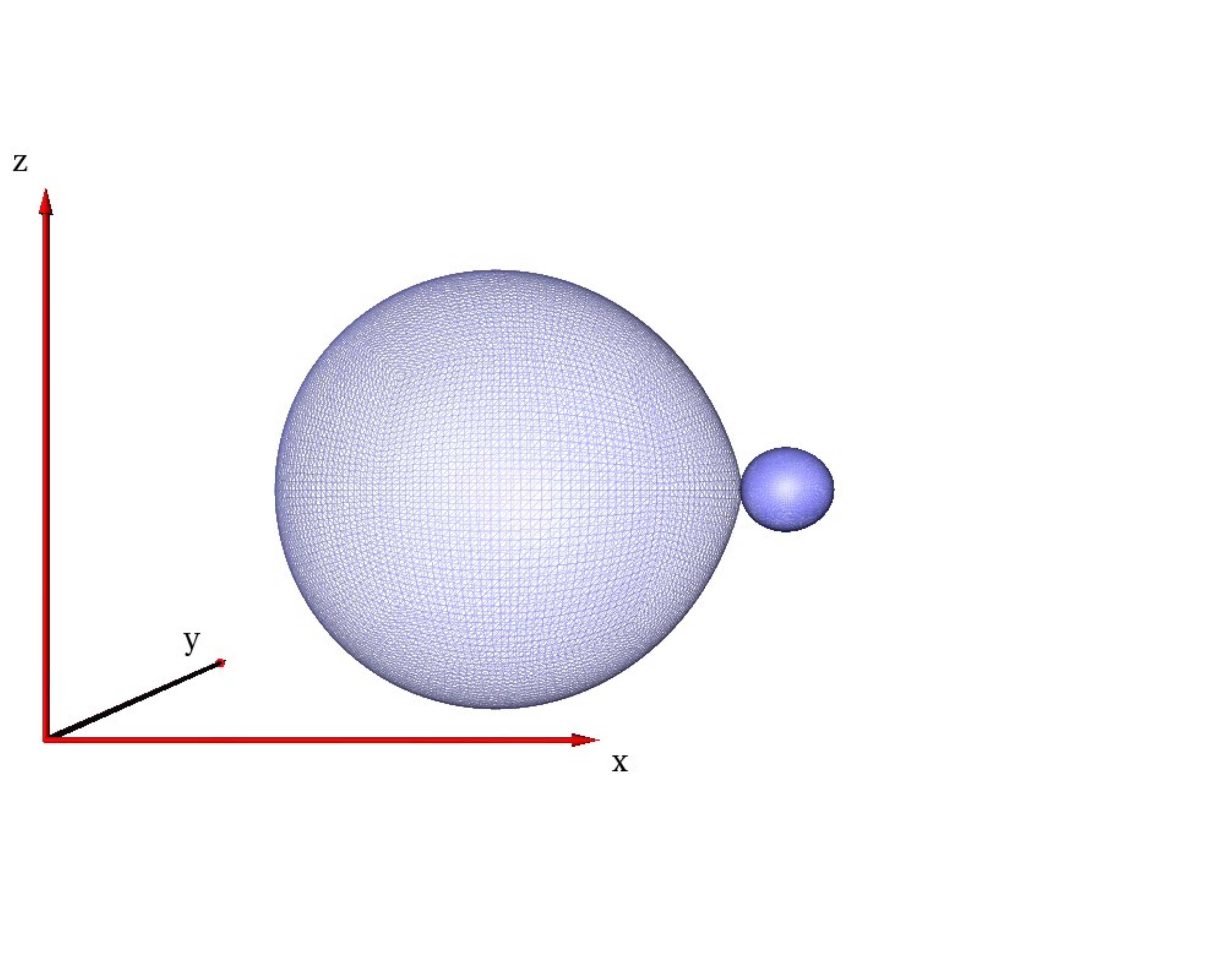}
    \includegraphics[width=.45\linewidth]{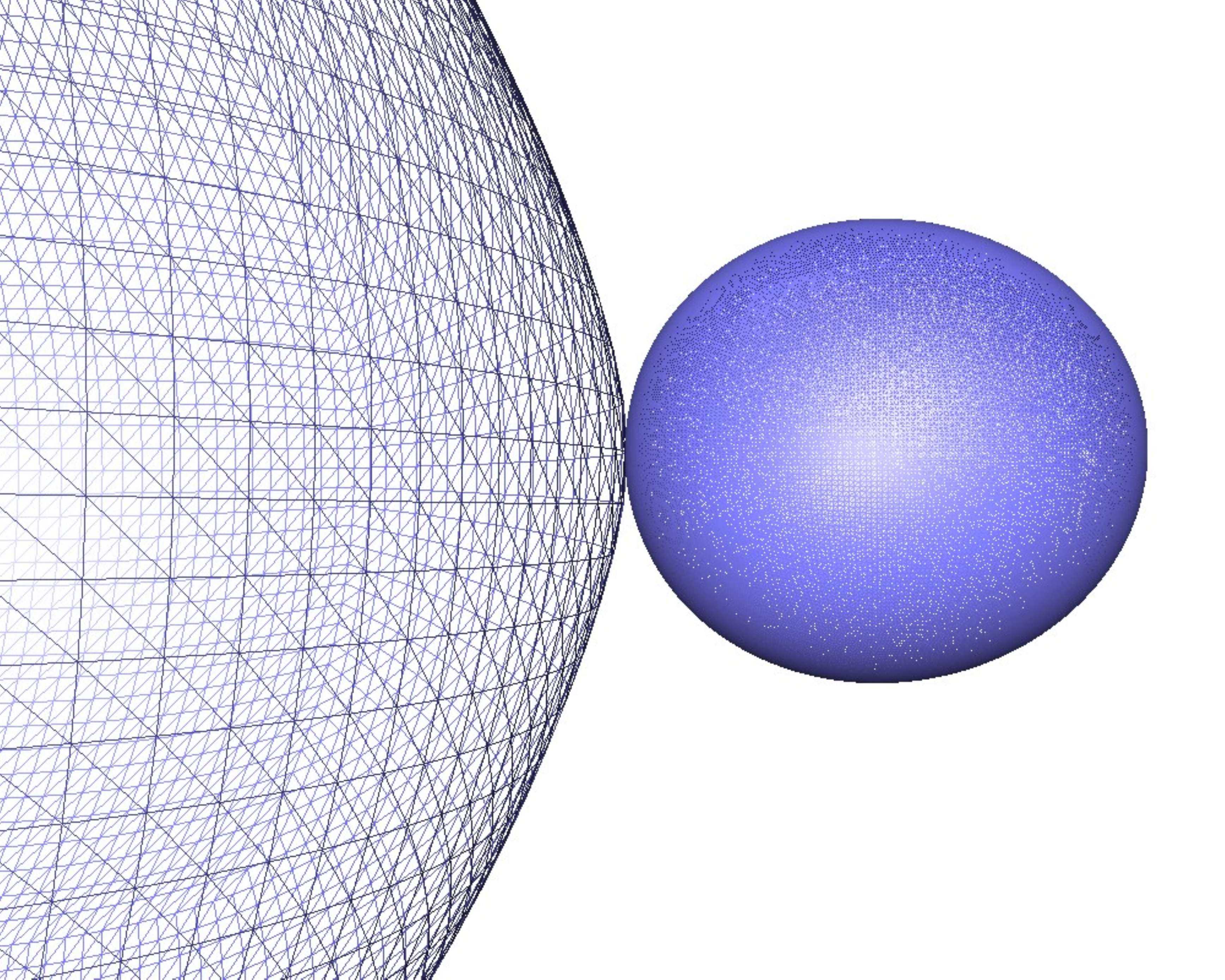}
  \end{center}
  \caption{Coordinate shapes of the individual MOTSs at time $t=1.97384M$
    when they touch. The picture on the right zooms in on the contact point,
    at which the two MOTS have equal mean curvature.}
  \label{fig:t3}
\end{figure}

\begin{figure}[!h]
  \begin{center}
    \includegraphics[width=.45\linewidth]{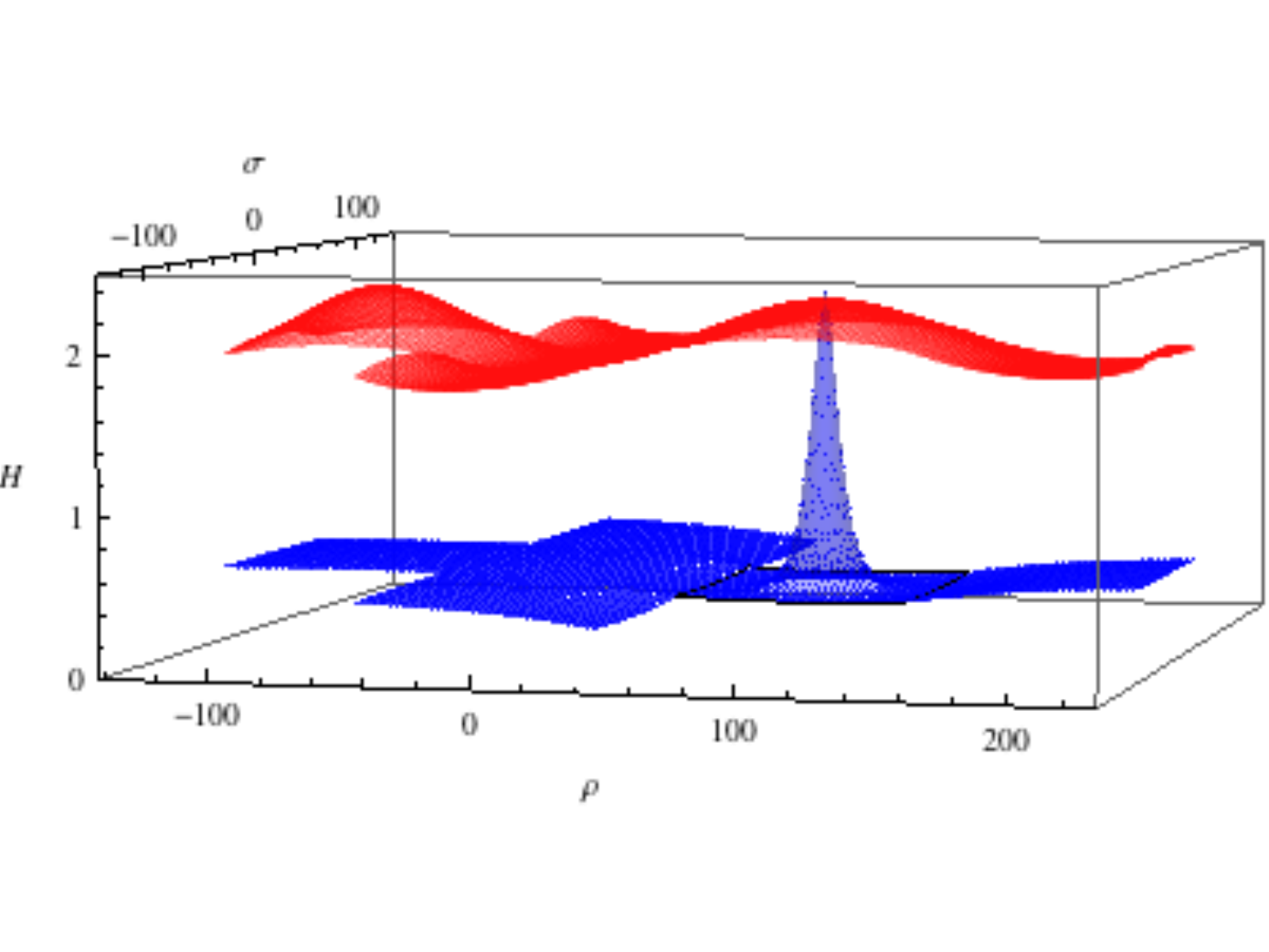}
  \end{center}
  \caption{Plots of the mean curvatures $H$ of the individual MOTSs laid out on
          the $(\rho,\sigma)$ coordinates of the
          six patches of the horizon finder
          at the time of osculation $t=1.97384M$. Their mean curvatures now agree at the
	contact point, as required by the theory. This is mainly achieved through the distortion
	of the large MOTS.}
  \label{fig:t3-2}
\end{figure}

The front-sides of the approaching MOTSs touch at $t=1.97384M$, as portrayed
by their coordinate representations in Fig.~\ref{fig:t3}. At this time, 
Fig.~\ref{fig:t3-2} shows surface plots of the mean curvatures of 
the individual MOTSs. Now the distortion has increased so that 
their mean curvatures are equal at their common point, as required 
by (\ref{eq:Hsml2}).

\subsection{Penetrating MOTSs}

\begin{figure}
  \begin{center}
    \includegraphics[width=.45\linewidth]{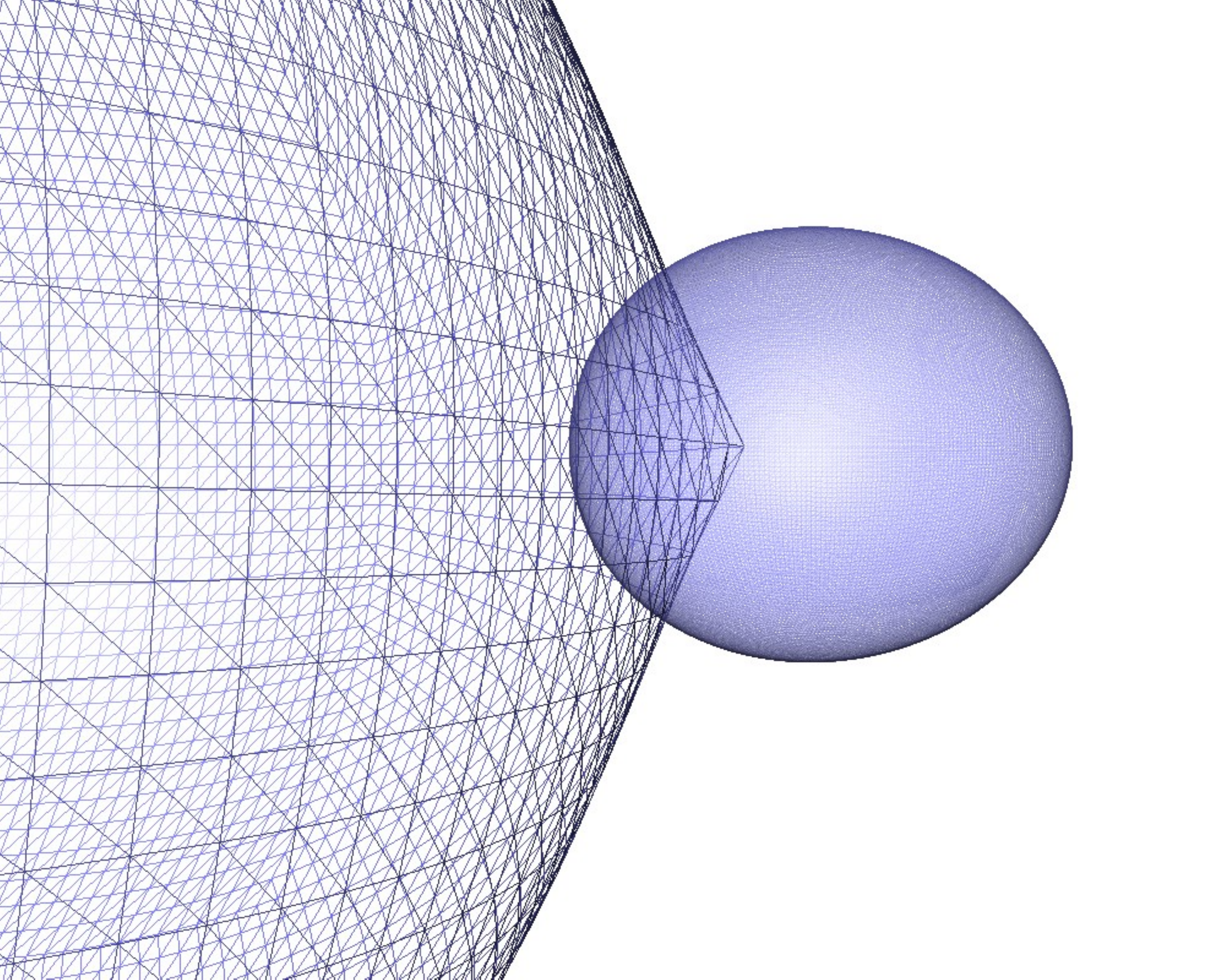}
    \includegraphics[width=.45\linewidth]{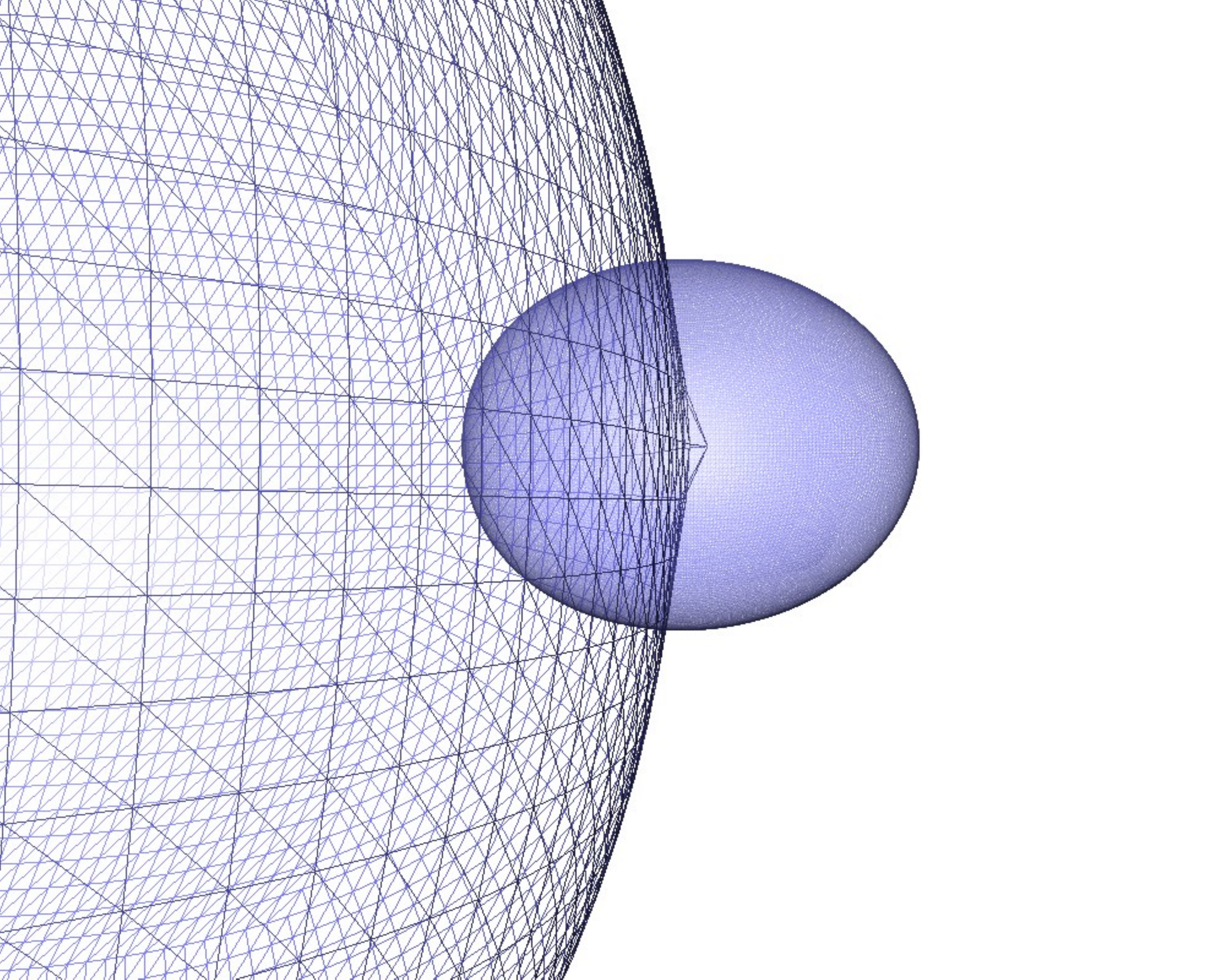}
  \end{center}
  \caption{Zoomed-in view of the coordinate shapes of both individual
MOTSs during penetration at times $t=2.1785M$ (left) and $t=2.538M$ (right).
The puncture near the center of the small MOTS has produced a  sharp effect on the front
end of the large MOTS.}
  \label{fig:t4}
\end{figure}

After osculation, the individual MOTSs continue to penetrate.
Figure \ref{fig:t4} shows the coordinate shapes of
both MOTSs as the penetration proceeds. The left panel of
Fig.~\ref{fig:t4} shows the two individual MOTSs at $t=2.1785M$. 
The coordinate shape of the larger MOTS is clearly
being affected along the axis where the small one has penetrated.
The right panel of Fig. \ref{fig:t4} illustrates 
the coordinate shapes at a later time $t=2.538M$, when the
small MOTS has penetrated approximately half-way inside the larger one.
At this stage, the puncture near the center of the small MOTS can be expected
to have a significant effect on the larger one. This is evident
in Fig.~\ref{fig:t4} from the sharp localized feature that has developed at the front of the
large MOTS.

As the simulation progresses further, the horizon finder is not
able to locate the larger individual MOTS.
This could be due either
to the horizon finder breaking down for numerical reasons
due to the puncture or simply due to 
the larger MOTS ceasing to exist. Further code development would
be necessary to resolve this issue.
 
\begin{figure}
  \begin{center}
    \includegraphics[width=.45\linewidth]{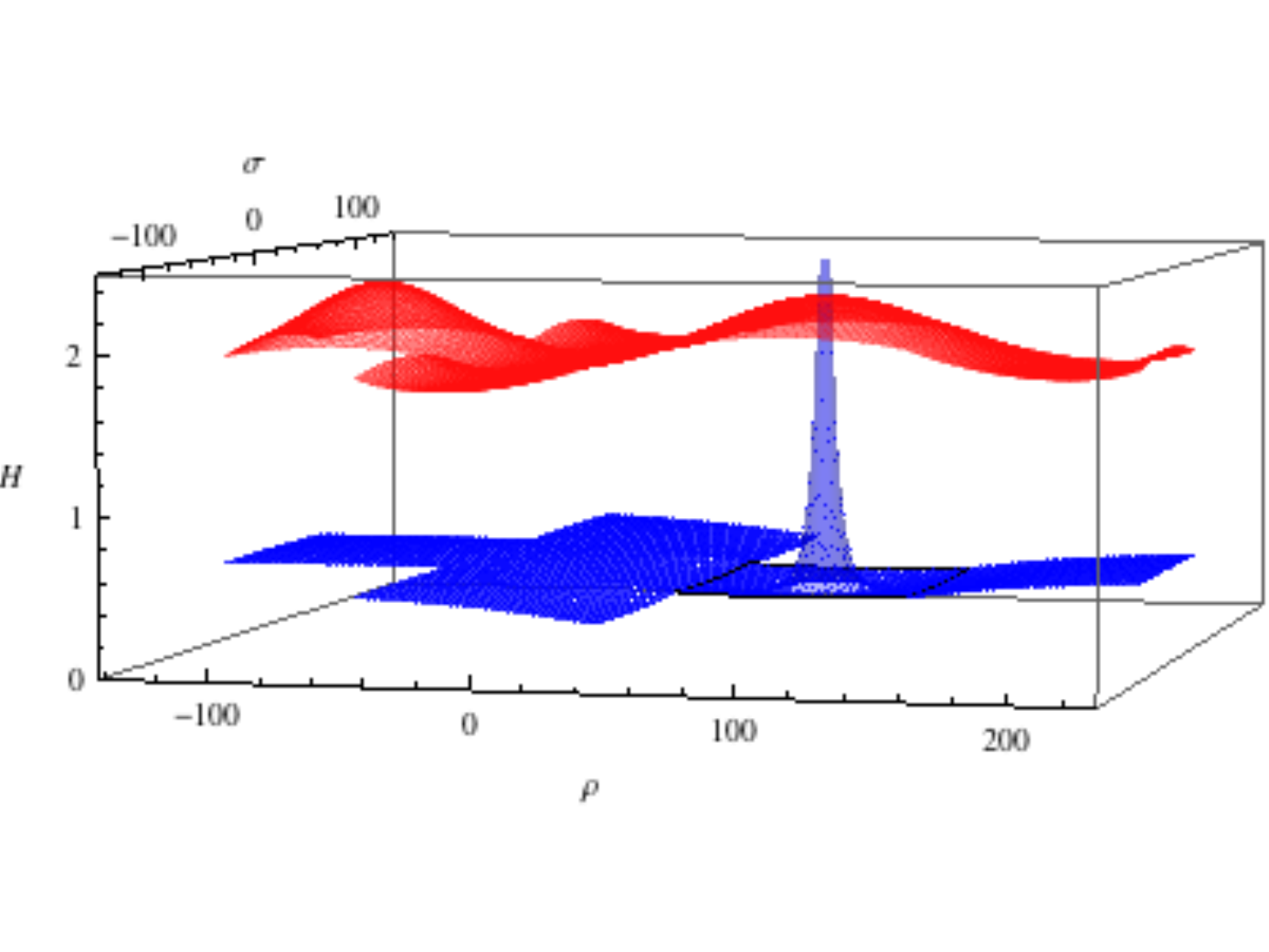}
    \includegraphics[width=.45\linewidth]{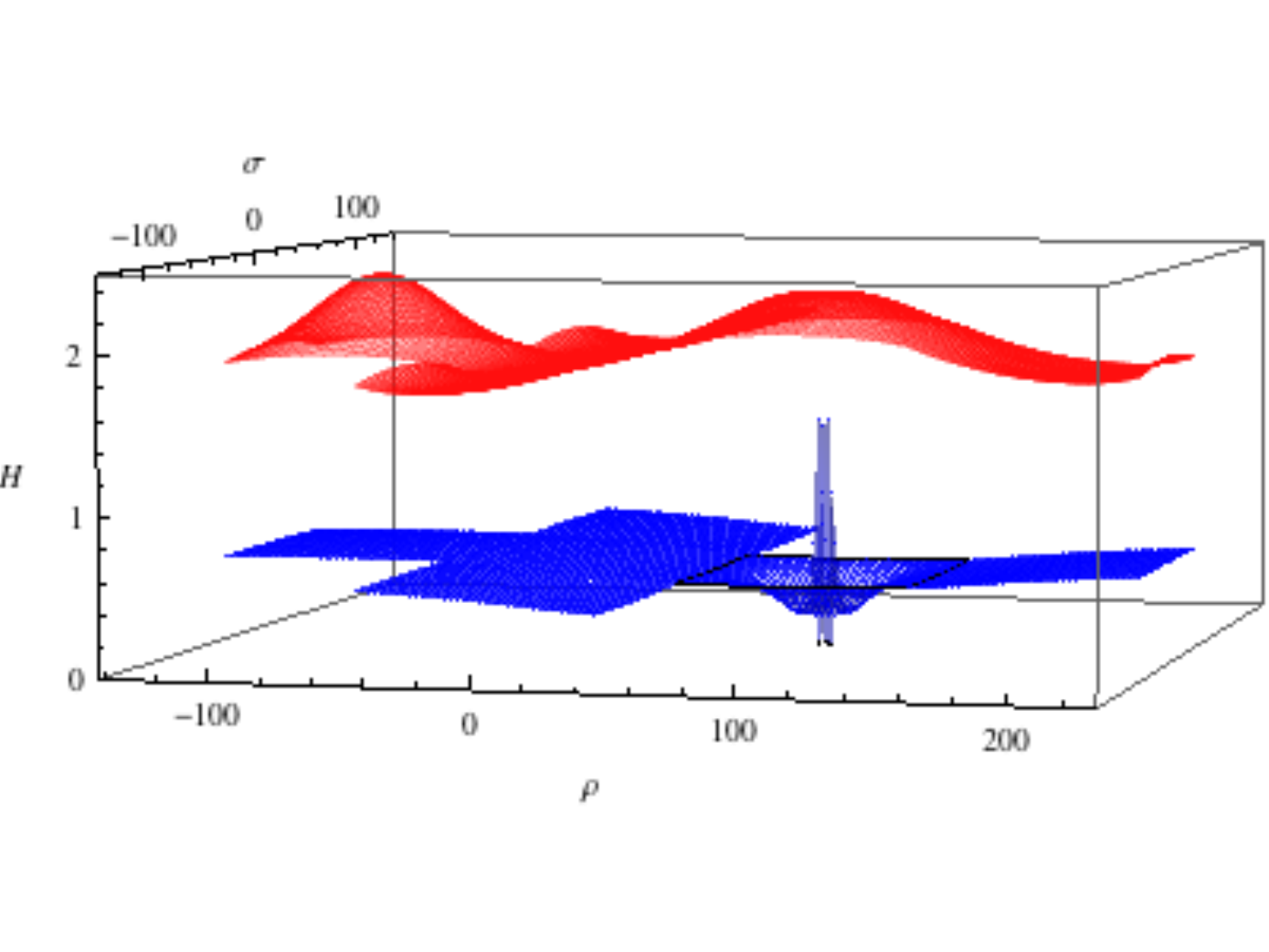}
  \end{center}
  \caption{Mean curvatures $H$ of the individual MOTSs laid out on the
	 $(\rho,\sigma)$ coordinates of the of the horizon finder
          at times $t=2.1785M$ (left) and $t=2.538M$ (right). On the left,
	 as the penetration first proceeds,
	the mean curvature at the front of the larger MOTS rises above that
         of the smaller MOTS. Later, on the right, the puncture of the smaller
         MOTS has had an oscillatory effect on the the mean curvature of the larger
	one. Comparing the two panels, the mean curvature of the smaller MOTS has
	undergone little change. Further evolution becomes problematic
	beyond this stage. }
  \label{fig:t5}
\end{figure}

The left and right panels of Fig.~\ref{fig:t5} give
surface plots of the mean-curvature of the individual MOTSs
at the same times $t=2.1785M$ and $t=2.538M$ as in Fig.~\ref{fig:t4}.
The left panel of Fig.~\ref{fig:t5} shows that the mean curvature
at the front of the larger MOTS has continued to grow and now
overshoots the mean curvature of the smaller MOTS. At the
later time displayed in the right panel of Fig.~\ref{fig:t5},
the mean curvature at the front of the larger MOTS
has now fallen back below the mean curvature of the smaller MOTS and has
developed some highly oscillatory behavior.
Again, this could be the result of the puncture inside the smaller
MOTS as it crosses the surface of the larger MOTS at $t\approx 2.45M$.

\begin{figure}
  \begin{center}
    \includegraphics[width=.45\linewidth]{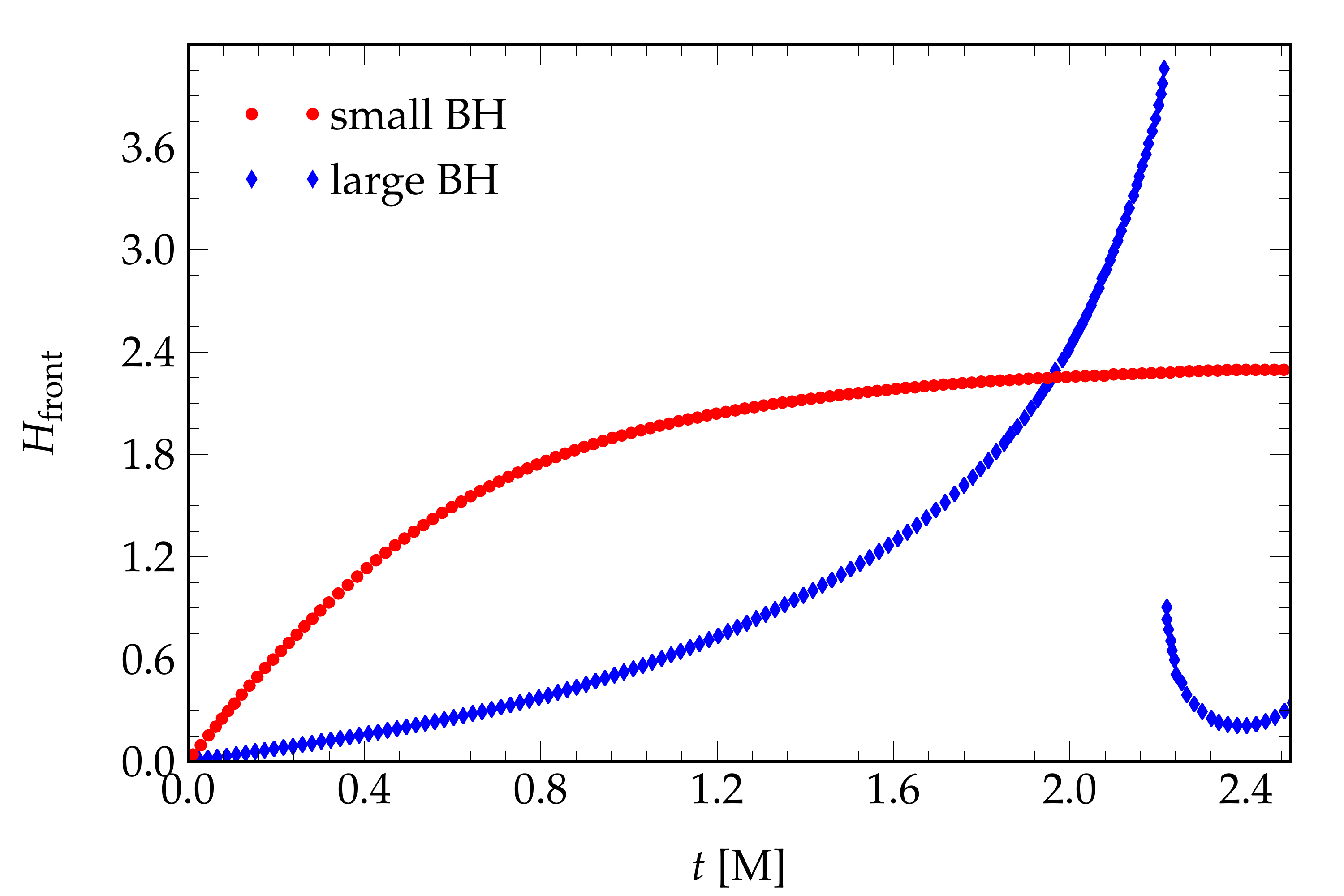}
  \end{center}
  \caption{Mean curvature $H$ at the front ends of both MOTSs
	as a function of time. $H_{(small)}$ grows steadily from its
	initial time symmetric value and then
	settles to a fairly constant value.
	$H_{(large)}$ rises rapidly as the MOTSs osculate and
 	penetrate and then drops discontinuously after the puncture
	inside the small MOTS passes through.}
  \label{fig:t5-3}
\end{figure}

Figure~\ref{fig:t5-3} shows the time evolution of the mean curvatures at the
(facing) front ends of the individual MOTSs as they approach and penetrate.
These are the points around which their mutual distortion is greatest.
The mean curvature at the front end 
of the smaller MOTS grows steadily from its
initial time symmetric value $H=0$ and then
settles to a fairly constant value as its geometry
adjusts to the ambient gauge.
In contrast, the mean curvature $H_{(large)}$ at the front end of the larger MOTS
shows a rapid increase as the smaller MOTS approaches and penetrates.
At the time of osculation, $t=1.97384M$, both mean curvatures
agree, as required by (\ref{eq:Hsml2}).
As the penetration continues, $H_{(large)}$ continues to rise until
the puncture inside the small MOTS crosses it.  $H_{(large)}$ 
then first drops discontinuously to a much smaller
value and then begins to rise again during the period when both
MOTS overlap. Runs
with different spatial resolutions and different time steps
confirm that this
jump is not a numerical artifact of the evolution code.
It could possibly result from a discontinuity in the
MOTT trace out by the large MOTS similar to the jump
that occurs in the formation of the outer common horizon.
Or it could be an indication of a more complicated horizon
structure which is difficult for the horizon finder to
fully describe. This is another issue that we defer to future
work.

\begin{figure}
  \begin{center}
    \includegraphics[width=.45\linewidth]{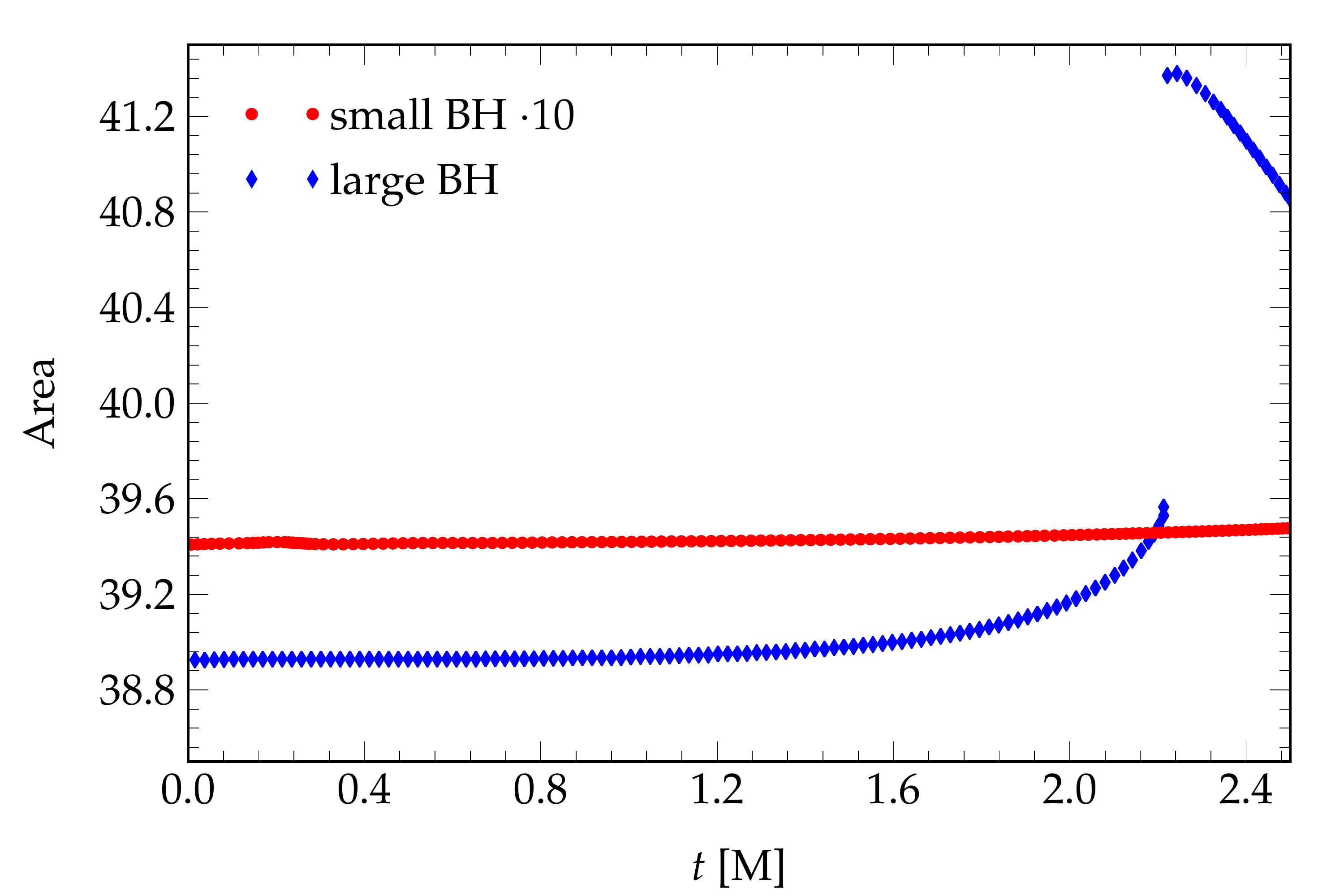}
  \end{center}
  \caption{Areas A of the small and large MOTSs as a function of time.
	 $A_{(large)}$ undergoes a discontinuous increase
	at the same time that its mean curvature drops  discontinuously.
	Afterwards,  $A_{(large)}$ drops continuously but at a rapid rate.
	 $A_{(small)}$, whose value is multiplied by 10 to match the scale of the plot,
	 shows very little variation
	during the entire run, which offers no clue regarding the possible future coalescence
	of the two MOTSs.}
  \label{fig:t5-4}
\end{figure}

This discontinuous drop in $H_{(large)}$ has a counterpart in 
Fig. \ref{fig:t5-4}, which plots the time evolution of
the areas $A$ of both individual MOTSs.
The larger MOTS shows
a discontinuous jump in area at the same time its mean curvature
undergoes a discontinuous drop.   
The small MOTS shows very small growth in area
during the entire run, with an approximate range
$3.941<A_{(small)}<3.948$. (Near the outset, $A_{(small)}$ undergoes
a small anomalous decrease which can be attributed to numerical error
since this would otherwise be inconsistent with the strict stability
of the small MOTS at this time.) At the end of the run, the ratio
$A_{(large)}/A_{(small)}$ has not undergone substantial change, which
offers no clue regarding their possible future coalescence.

\subsection{Convergence}
\label{sec:convergence}

In order to verify the new geometrical features we have found,
we have measured the convergence of the
harmonic constraints ${\cal C}^\mu$, i.e. (\ref{eq:hc}) with the gauging
forcing $F^\mu=0$ used in our simulations. We monitor the constraints
via the norm
\begin{equation}
         |{\cal C}|^2:= e_{\mu\nu}C^{\mu}C^{\nu},
\end{equation}
where $e_{\mu\nu}$ is the positive-definite metric defined in (\ref{eq:reimmet}).
Although the code has
fourth order finite difference accuracy, the time integrator used in 
conjunction with the adaptive mesh refinement limits the overall 
accuracy to second order. Confirmation of second order convergence 
of $ |{\cal C}|$ in a region including the penetration of the MOTSs supplies the 
necessary code verification.

\begin{figure}[!h]
  \begin{center}
    \includegraphics[width=.45\linewidth]{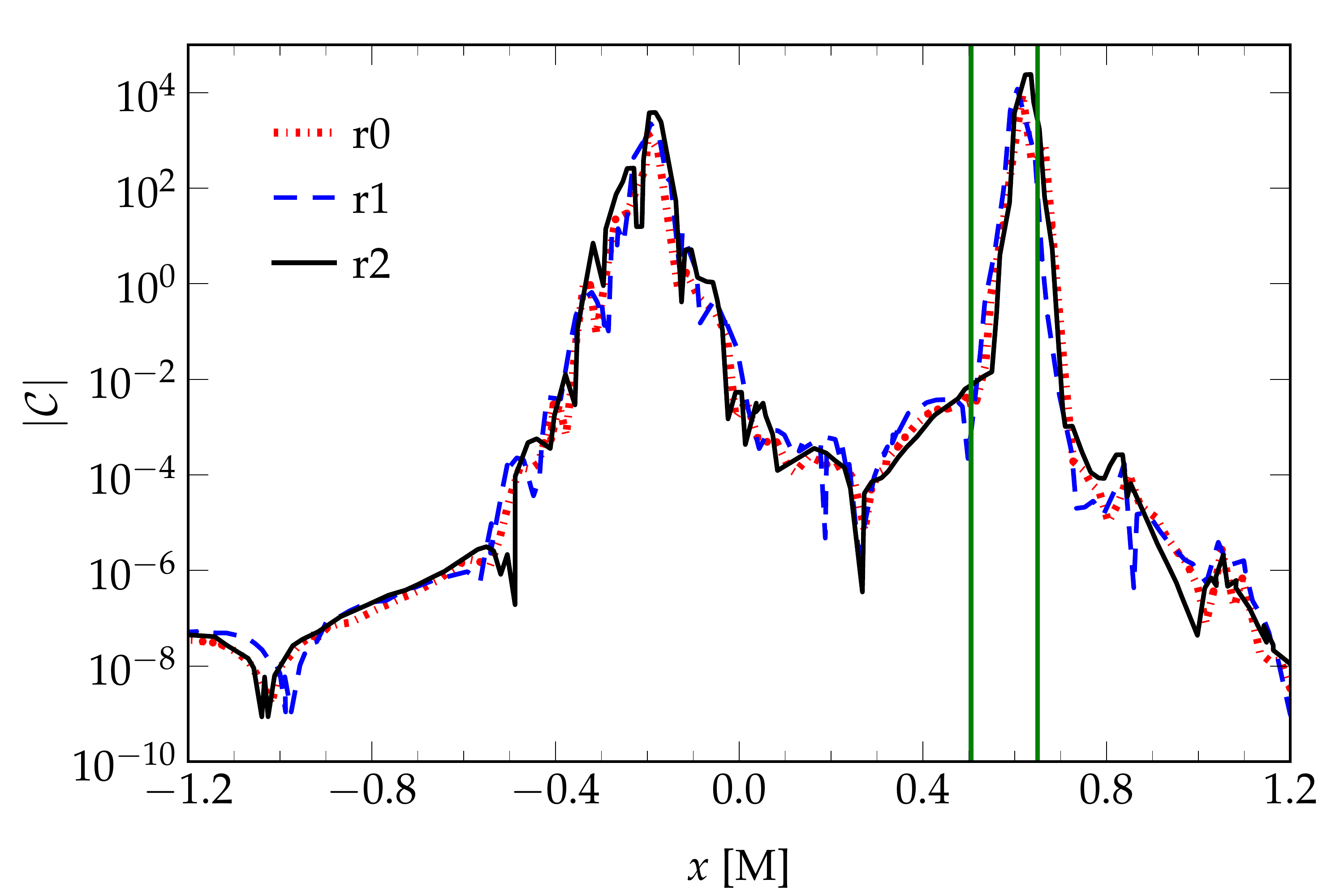}
  \end{center}
  \caption{Rescaled norms of the harmonic constraints $|{\cal C}|$ along
	the x-axis for 3 different resolutions $(r0,r1,r2)$,
	corresponding to the gridsizes $h=(3.75\cdot10^{-3}M,4.375\cdot10^{-3}M,5\cdot10^{-3}M)$,
	at time $t=2.45M$.
	The norms have been rescaled for resolution to confirm that the convergence is
	second order. The vertical lines bound the region that the penetrating MOTS
	overlap.}
  \label{fig:t6}
\end{figure}

Figure \ref{fig:t6} plots $|{\cal C}|$ along the $x$-axis,
in the relevant region where the puncture and
outer boundary are not included, for three different gridsizes,
 $h=(3.75\cdot10^{-3}M,4.375\cdot10^{-3}M,5\cdot10^{-3}M)$,
at time $t=2.45M$. At this time, the individual MOTSs have penetrated
and overlap in the region between $x=0.45M$ and $x=0.63M$, as indicated
by the vertical lines in the figure marking the front and back end of both MOTSs.
The curves are rescaled in accordance with the expected second order convergence
of the constraints. Second-order convergence is confirmed
even very close to the punctures in the region where
the MOTSs overlap. This confirms the validity of the
geometrical features we have observed of the MOTSs and their penetration.

\section{Conclusion}

The results presented here give independent confirmation of the penetration
of individual MOTSs reported in~\cite{pen} and, by using punctures rather
than excision, we are able to follow the penetration to approximately
the halfway stage.
While the current code also works well with the excision module
developed for the PITT Abigel code (and later
adapted to the AEI Harmonic code~\cite{pen}),
the excised domain restricts the simulation to a very small penetration.
The horizon finder algorithm would fail as soon as 
its required search domain intersected an excised region.
Unfortunately, even with the use of punctures, the tracking of the larger MOTS breaks down before
the most interesting final stage. Either the individual MOTSs eventually cease to
exist or hit the final singularity; or otherwise they must coalesce. 

The apparently discontinuous behavior of the large MOTS near the end of the simulation
indicates that there might be underlying complexities introduced by the time slicing.
It is particularly interesting that the jump leads first to an increase in its area, followed
by a continuous fall. In
order to investigate this feature and track the penetration further, it might
be necessary to redesign
the horizon finder or to make some
fortuitous choice of harmonic gauge source function. This is an open problem
since this is the first binary black hole simulation carried out in the harmonic
formulation using punctures. 

One attractive approach to the
code modifications necessary to continue our simulations
is to fill the puncture with artificial matter, using a version of the
``turduckening'' procedure~\cite{turduck}.  This is feasible because
the uniqueness theorem governing the coalescence 
is independent of a positive energy
condition on the matter.

The penetration of MOTSs is a new aspect of the
of the dynamics of MOTTs. The possibility exists that the individual
MOTTs merge and connect
back to the  (unstable) branch of the common outer horizon to form a globally
connected MOTT, This makes the system
of penetrating MOTSs a rich laboratory for the geometry governed by Einstein's equations.

\begin{acknowledgments}

We have benefited from discussions with B. Krishnan and E. Schnetter.
JW's research was supported by NSF grant PHY-1201276 to the University of Pittsburgh.
PM and BS were supported by the Sherman Fairchild Foundation
and NSF grants PHY-1440083, AST-1333520 and AST-1212170 at Caltech.

\end{acknowledgments}

\end{document}